\documentclass[prb,aps,10pt,twocolumn,twoside,superscriptaddress,showkeys]{revtex4-2}

\usepackage{amsmath,amsfonts,amssymb,amsthm,mathtools}

\usepackage[english]{babel}
\usepackage[utf8]{inputenc}
\usepackage[T1]{fontenc}

\usepackage{latexsym}
\usepackage[pdftex]{graphicx}
\usepackage{epstopdf}

\usepackage{subfigure}
\usepackage{enumitem}

\usepackage{hyperref}
\usepackage[all]{hypcap}

\hypersetup{
	pdftitle={},
	pdfauthor={},
	colorlinks=true,
	linkcolor=black,
	citecolor=black,
	filecolor=black,
	urlcolor=blue,
	bookmarksopen,
	bookmarksopenlevel=1,
}

\newcommand{\cE}{\mathcal{E}}
\newcommand{\cF}{\mathcal{F}}

\newcommand{\cO}{\mathcal{O}}

\newcommand{\ee}{\mathrm{e}}
\newcommand{\ii}{\mathrm{i}}
\newcommand{\dd}{\mathrm{d}}

\def\io{\infty}
\def\eps{\epsilon}

\def\tx{x}
\def\ty{y}

\def\Tr{\operatorname{Tr}}

\def\wDiff{\widetilde{\operatorname{Diff}}}

\newcommand\Tstrut{\rule{0pt}{2.6ex}}				

\theoremstyle{definition}

\theoremstyle{remark}

\begin{document}


\title{%
Geometric approach to inhomogeneous Floquet systems%
}%

\author{Bastien Lapierre}
\email{blapie@physik.uzh.ch}
\affiliation{%
Department of Physics, University of Zurich,
Winterthurerstrasse 190, 8057 Z{\"u}rich, Switzerland%
}%

\author{Per Moosavi}
\email{pmoosavi@phys.ethz.ch}
\affiliation{%
Institute for Theoretical Physics, ETH Zurich,
Wolfgang-Pauli-Strasse 27, 8093 Z{\"u}rich, Switzerland%
}%

\date{%
June 8, 2021%
}%

\begin{abstract}
We present a new geometric approach to Floquet many-body systems described by inhomogeneous conformal field theory in 1+1 dimensions. It is based on an exact correspondence with dynamical systems on the circle that we establish and use to prove existence of (non)heating phases characterized by the (absence) presence of fixed or higher-periodic points of coordinate transformations encoding the time evolution: Heating corresponds to energy and excitations concentrating exponentially fast at unstable such points while nonheating to pseudoperiodic motion. We show that the heating rate (serving as the order parameter for transitions between these two) can have cusps, even within the overall heating phase, and that there is a rich structure of phase diagrams with different heating phases distinguishable through kinks in the entanglement entropy, reminiscent of Lifshitz phase transitions. Our geometric approach generalizes previous results for a subfamily of similar systems that used only the $\mathfrak{sl}(2)$ algebra to general smooth deformations that require the full infinite-dimensional Virasoro algebra, and we argue that it has wider applicability, even beyond conformal field theory.
\end{abstract}


\maketitle


\section{Introduction}
\label{Sec:Introduction}


Floquet drives (also known as periodic drives) in quantum many-body systems
are well-known mechanisms for creating nonequilibrium states of matter, such as Floquet topological insulators \cite{LRG:2011, RechEtAl:2013, TBRRL:2016, NRLBR:2017, StutzerEtAl:2018} and time crystals \cite{ChoiEtAl:2017, ZhangEtAl:2017}.
They also provide a fruitful setting to study a range of active problems in physics, including nonequilibrium topological properties \cite{JiangEtAl:2011, KBRD:2010, RLBL:2013, GrTa:2018}, many-body-localization transitions \cite{PPHA:2015, ARH:2016, KLMS:2016, ZKH:2016}, prethermalization \cite{ARH:2015, BHHP:2016, ARHH1:2017, ARHH2:2017}, and driven lattice vibrations \cite{RTDITTSC:2007, FCKTTMC:2011, MSFMCLRGMFFSLKGC:2014, LFFNPLKC:2020}.
However, despite these recent advances, Floquet many-body systems remain challenging to study theoretically by exact analytical means.

One framework amenable to exact analytical solutions is given by conformal field theory (CFT) in 1+1 dimensions.
Here, we focus on the recent concept of inhomogeneous CFT, whose applications include effective descriptions of trapped ultracold atoms, inhomogeneous gapless spins chains, and quantum fluctuations in generalized hydrodynamics \cite{ADSV:2016, DSVC:2016, DSC:2016, GLM:2018, LaMo2:2019, Moo1:2019, MLNR:2019, RCDD:2020}.
An example of such a Floquet system is a two-step drive alternating between two different CFTs.
This was recently studied for the special case of homogeneous and sine-square-deformed (SSD) CFTs using M{\"o}bius transformations \cite{WeWu2:2018, LCTTNC1:2020, FGVW:2020, HaWe:2020}.
Yet, general deformations have not been considered until now.
This is an important extension since generic deformations would be more realistic for experiments (as engineering too specific deformations is likely beyond experimental control).
To do this, more powerful tools are needed, which we develop here based on \cite{GLM:2018, Moo1:2019, GaKo:2020}.
Our main result is a new geometric approach to Floquet systems, which allows one to generate a rich structure of phase diagrams, among others exhibiting new nonequilibrium analogues of Lifshitz phase transitions \cite{RSLLS:2013}, and we argue that this approach and the underlying ideas are relevant even beyond CFT.

\begin{figure}[!htbp]

\centering

\includegraphics[scale=1, trim=0 0 0 0, clip=true]
{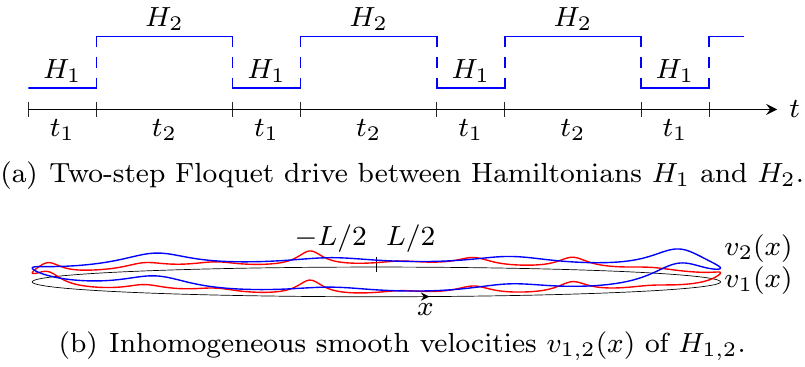}%

\vspace{-2.0mm}

\caption{%
Illustrations for our inhomogeneous Floquet drive.%
}

\label{Fig:Floquet_drive}

\end{figure}

In this paper, for simplicity, we present our approach by using it to study two-step Floquet drives for general smoothly deformed CFTs with periodic boundary conditions.
The time evolution is under a Hamiltonian that equals $H_{1}$ or $H_{2}$ for times $t_{1}, t_{2} \in \mathbb{R}$ \cite{Note0}, respectively, see Fig.~\ref{Fig:Floquet_drive}(a).
The associated Floquet operator is
\begin{equation}
\label{U_F}
U_{F}
= \ee^{-\ii H_{1} t_{1}} \ee^{-\ii H_{2} t_{2}},
\end{equation}
and the period for a full cycle is $t_{\mathrm{cyc}} = |t_{1}| + |t_{2}|$.
Here, $H_{1}$ and $H_{2}$ are given by any smooth deformation of standard Minkowskian CFT on the cylinder $\mathbb{R} \times [-L/2, L/2]$.

We refer to the above general smoothly deformed CFT as \emph{inhomogeneous CFT} \cite{GLM:2018, LaMo2:2019, Moo1:2019}.
They are defined by
\begin{equation}
\label{H_1_2}
H_{1, 2}
= \int_{-L/2}^{L/2} \dd x\, v_{1, 2}(x) \bigl[ T_{+}(x) + T_{-}(x) \bigr]
\end{equation}
for smooth velocity functions $v_{1, 2}(x) = v_{1, 2}(x+L) > 0$, see Fig.~\ref{Fig:Floquet_drive}(b), where $T_{\pm}(x) = T_{\pm}(x+L)$ are the usual light-cone components of the energy-momentum tensor of a two-dimensional Minkowskian CFT.
These satisfy
\begin{align}
\bigl[ T_{\pm}(x), T_{\pm}(y) \bigr]
= & \mp 2\ii \delta'(x-y) T_{\pm}(y)
		\pm \ii \delta(x-y) T_{\pm}'(y) \nonumber \\
	& \pm ({c}/{24\pi}) \ii \delta'''(x-y)
\end{align}
and $\bigl[ T_{\pm}(x), T_{\mp}(y) \bigr] = 0$, where $c$ is the central charge of the CFT and the subscript $+$($-$) denotes the right- (left-) moving component.
These commutation relations correspond to two commuting copies of the Virasoro algebra \cite{Note1}.
We will use this full infinite-dimensional algebra, which generalizes previous works on Floquet drives in deformed CFTs that involved only its finite-dimensional $\mathfrak{sl}(2)$ subalgebra, cf.\ also \cite{GrPo:2017}.

\begin{figure}[!htbp]

\centering

\includegraphics[scale=1, trim=0 0 0 0, clip=true]
{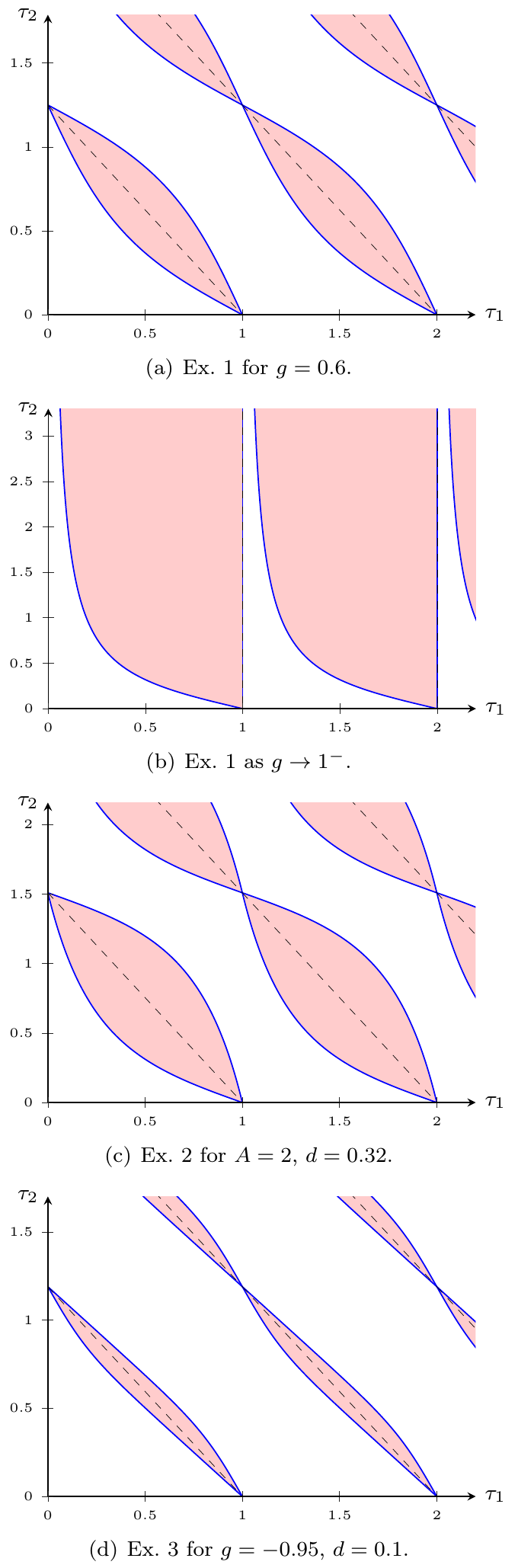}%

\vspace{-2.0mm}

\caption{%
Phase diagrams in $(\tau_{1}, \tau_{2})$-space for $H_{1}$ with $v_{1}(x) = v_{1}$ and $H_{2}$ with $v_{2}(x) = v_{2} w_{2}(x/L)$ in Table~\ref{Table:Examples}:
(a,~b) Ex.~1,
(c) Ex.~2, and
(d) Ex.~3.
Blue lines signify transitions between nonheating (white regions) and heating (pink regions) \cite{Note2}, obtained by solving \eqref{p-periodic_points} and \eqref{tangent_points} analytically.
Black dashed lines are given by
$\tau_{2} = m/k_{1}w_{2, 0} - (k_{2}w_{1, 0}/k_{1}w_{2, 0}) \tau_{1}$, $m \in \mathbb{Z}$.%
}

\label{Fig:Phase_diagrams}

\end{figure}

The deformations in our inhomogeneous CFTs are contained in $v_{1, 2}(x)$, with constant velocity corresponding to standard homogeneous CFT.
It is convenient to write $v_{1, 2}(x) = v_{1, 2} w_{1, 2}(x/L)$ for constants $v_{1, 2} > 0$ and dimensionless functions $w_{1, 2}(\xi)$ of $\xi = x/L$, see Table~\ref{Table:Examples} for a selection of examples.
We also introduce dimensionless times $\tau_{1, 2} = v_{1, 2} t_{1, 2}/L$ as the true independent parameters used to plot phase diagrams, see Fig.~\ref{Fig:Phase_diagrams}.
The general structure of such phase diagrams is as follows:
There are \emph{leaf-shaped regions} where the system heats up, characterized by fixed or higher-periodic points of coordinate transformations encoding the time evolution, surrounded by regions where the system does not heat up.
Moreover, the diagrams are invariant under the following:
\begin{enumerate}[labelindent=1em, nosep, label={\textnormal{\arabic*.}}, ref={\textnormal{\arabic*.}}]

\item
\label{Item:Time_reversal_symmetry}
changing $(\tau_{1}, \tau_{2})$ to $(-\tau_{1}, -\tau_{2})$,

\item
\label{Item:Translation_symmetry_1}
translations in $\tau_{1}$ by $1/k_{2}w_{1, 0}$,

\item
\label{Item:Translation_symmetry_2}
translations in $\tau_{2}$ by $1/k_{1}w_{2, 0}$,

\end{enumerate}
where $w_{j, 0}^{-1}$ is given by
$w_{j, 0}^{-1} = \int_{-1/2}^{1/2} \dd \xi\, w_{j}(\xi)^{-1}$
and $k_{j}^{-1}$ is the period of $w_{j}(\xi)$ for $j = 1, 2$.
[If $w_{j}(\xi)$ is constant, we set $k_{j} = 1$.]
The above follow from general properties for our periodic points (see Sec.~\ref{Sec:Main_tools:Properties_pps}) that we will prove.

\begin{table}[!htbp]

\begin{tabular}{c|c|c}
\hline
\hline
Example & $w(\xi)$ & Parameters \\
\hline
\Tstrut
Ex.~1
	& $1 + g [2\cos^2(\pi \xi) - 1]$
	& $g \in [0, 1)$ \\
Ex.~2
	& $A \ee^{-(\xi/d)^2}$
	& $A, d \in \mathbb{R}^{+}$ \\
Ex.~3
	& $1/[ 1 - g \cosh(\xi/d)^{-2} ]$
	& $g \in (-1, 1)$, $d \in \mathbb{R}^{+}$ \\
Ex.~4
	& $a/[ b + \sin(2\pi k \xi) + \cos(2\pi \xi) ]$
	& $a > 0$, $b > 2$, $k \in \mathbb{Z}^{+}$ \\
\hline
\hline
\end{tabular}

\caption{%
Examples of deformations $w(\xi)$ for $\xi = x/L$, including
SSD CFT (obtained from Ex.~1 as $g \to 1^{-}$) and by a Gaussian (Ex.~2).
Periodicity is imposed at $\xi = \pm 1/2$ if not manifest.%
}

\label{Table:Examples}

\end{table}

The geometric approach we propose is precisely the above-mentioned reduction to studying coordinate transformations.
Here, these transformations will be shown to be given by orientation-preserving diffeomorphisms of the circle obtained directly from the velocities $v_{1, 2}(x)$.
We stress that the choice of a two-step drive is for concreteness and simplicity:
Our geometric approach is straightforwardly applicable to more general drives, including multistep, random, chaotic, and quasiperiodic ones, cf.\ \cite{LCTTNC2:2020, WFVG1:2020, ABI:2020}.
Indeed, how to use our tools to study random deformations was recently demonstrated in \cite{LaMo2:2019}.
It can also be generalized to several coupled or uncoupled CFTs with different deformations and drives (the case here is equivalent to two uncoupled chiral CFTs with the same drive).

As is well known, an inescapable issue is that a CFT description of a given condensed matter system, in general, starts to lose validity as it heats up since the system exits the low-temperature regime.
Nonetheless, we expect that our approach and underlying ideas have wider applicability, even beyond CFT.
For instance, periodic points in the CFT description may prove to be robust to higher excitations, or the latter may be taken into account by other means.
Our approach is also suitable for numerical implementations, and we hope that it can be adapted or serve as inspiration for a wide range of other Floquet systems, e.g., for driven Bose-Einstein condensates \cite{ChSh:2020, LLZ:2020, ZhGu:2020} or fields in modulated cavities \cite{Mart:2019}.

The rest of this paper is organized as follows.
In Sec.~\ref{Sec:Approach_and_Summary} we describe our approach and briefly summarize our results.
Our main tools are explained in Sec.~\ref{Sec:Main_tools} and used to lay the mathematical foundations for our approach.
In Sec.~\ref{Sec:cE_and_2pcf} we derive exact analytical results for the energy density and the flow of excitations.
The corresponding results for entanglement entropy and mutual information are given in Secs.~\ref{Sec:EE} and~\ref{Sec:MI}, respectively.
Lastly, Sec.~\ref{Sec:Concluding_remarks} contains concluding remarks.
Proofs of properties for the periodic points, which imply the symmetries of the phase diagrams illustrated in Fig.~\ref{Fig:Phase_diagrams}, and computational details for certain results are deferred to Appendices~\ref{App:Proof_of_properties_pps}--\ref{App:MI}.


\section{Approach and summary of results}
\label{Sec:Approach_and_Summary}


The key step is to establish an exact correspondence with dynamical systems on the circle, see, e.g., \cite{dMvS:1993}, in the sense that the Floquet time evolution can be fully encoded into coordinate transformations given by orientation-preserving diffeomorphisms of the circle, see Sec.~\ref{Sec:Main_tools}.
These diffeomorphisms $f_{\pm}(x)$ are obtained directly from the velocity functions $v_{1, 2}(x)$ [see \eqref{f_pm}].

We specifically show that the system can be either nonheating or heating, with heating phases characterized by periodic points $\tx_{\ast p}^{\mp}$ of $f_{\pm}$ with period $p \in \mathbb{Z}^{+}$:
\begin{equation}
\label{p-periodic_points}
\tx_{\ast p}^{\mp} = f_{\pm}^{p}(\tx_{\ast p}^{\mp}),
\qquad
f_{\pm}^{p} = f_{\pm} \circ \ldots \circ f_{\pm},
\end{equation}
where the function composition $\circ$ is repeated $p$ times.
Of these, $\tx_{\ast 1}^{\mp}$ are fixed points of $f_{\pm}$, while $\tx_{\ast p}^{\mp}$ for $p > 1$ can be viewed as fixed points of $f_{\pm}^{p}$ for the smallest possible $p$.
On general grounds, since the deformations are the same in both terms in \eqref{H_1_2}, $f_{\pm}$ have equally many periodic points with stable and unstable ones coming in pairs.
Mergers of such pairs correspond to critical values $\tx_{\mathrm{c} p}^{\mp} = \tx_{\ast p}^{\mp}$ denoting periodic points that also satisfy
\begin{equation}
\label{tangent_points}
1
= f_{\pm}^{p \,\prime}(\tx_{\mathrm{c} p}^{\mp})
= \frac{\dd}{\dd x} f_{\pm}^{p}(x)
  \Big|_{x = \tx_{\mathrm{c} p}^{\mp}}.
\end{equation}
Unstable points play a predominant role as their presence will imply that the system is heating up:
The picture is that energy and excitations flow to these points, which follows from exact analytical results for the evolution of the energy density and correlation functions, see Sec.~\ref{Sec:cE_and_2pcf}.

\begin{figure}[!htbp]

\centering

\includegraphics[scale=1, trim=0 0 0 0, clip=true]
{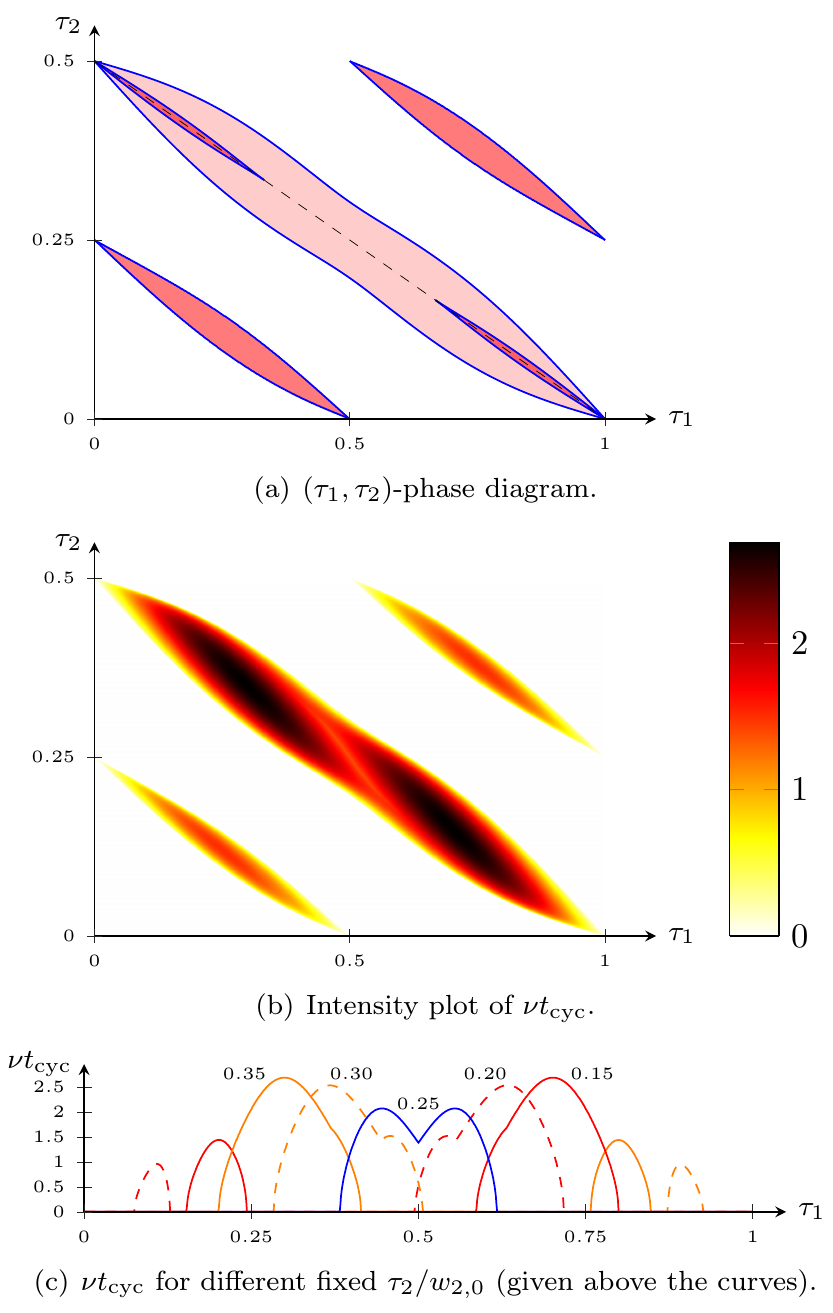}%

\vspace{-2.0mm}

\caption{%
Phase diagram and heating rate for $H_{1}$ given by homogeneous CFT and $H_{2}$ by Ex.~4 in Table~\ref{Table:Examples} with $a = 6$, $b = 3$, and $k = 2$.
(a)
Phase diagram in $(\tau_{1}, \tau_{2}) \in [0, 1] \times [0, 1/w_{2, 0}]$ (cf.\ Fig.~\ref{Fig:Phase_diagrams}) due to fixed points (center leaf) computed analytically and periodic points with period 2 (leaves on each side of the center) computed numerically.
Regions with two and four unstable periodic points are light and dark pink, respectively.
(There are regions due to periodic points with period 3 not shown, and possibly also due to ones with period $p > 3$.)
(b,~c)
Rescaled heating rate $\nu t_{\mathrm{cyc}}$ as function of $(\tau_{1}, \tau_{2})$ and along rays with fixed $\tau_{2}$, respectively.%
}

\label{Fig:SinCos_CFT_1}

\end{figure}

Let $2N_{p}$ be the number of periodic points $\tx_{\ast p, i}^{\mp}$ of $f_{\pm}$ with period $p$ ($i = 1, \ldots, 2N_{p}$).
We will show that the heating rate $\nu$ is given by \cite{Note3}
\begin{equation}
\label{heating_rate}
\nu
= \max_{p \in \mathbb{Z}^{+}\!,\, r = \pm,\, i \in \{1, \ldots, 2N_{p}\} }
	2 t_{\mathrm{cyc}}^{-1} \, p^{-1}
	\log[f_{r}^{p \,\prime}(\tx_{\ast p, i}^{-r})].
\end{equation}
If all $N_{p} = 0$, then $\nu = 0$, corresponding to nonheating, while $\nu > 0$ indicates heating.
Thus, knowledge of $f_{\pm}$ and \eqref{heating_rate} is all one needs to draw phase diagrams in $(\tau_{1}, \tau_{2})$-space, such as in Figs.~\ref{Fig:Phase_diagrams} and~\ref{Fig:SinCos_CFT_1}; see, e.g., Appendix~\ref{App:Practical_eqs} for how equation systems of the form in \eqref{p-periodic_points} and \eqref{tangent_points} can be solved in practice for phases due to fixed points.
We stress that Fig.~\ref{Fig:Phase_diagrams}(b) is obtained as $g \to 1^{-}$ for Ex.~1 in Table~\ref{Table:Examples} \cite{Note4} and agrees perfectly with Fig.~1(c) in \cite{LCTTNC1:2020}, which can be verified analytically, see Appendix~\ref{App:Analytical_results:Ex1}.

The heating rate serves as the natural order parameter for phase transitions between heating and nonheating.
However, we will also show that regions with different numbers of unstable periodic points can be distinguished by kinks in the entanglement entropy, reminiscent of Lifshitz phase transitions, see Sec.~\ref{Sec:EE}.
To better understand the entanglement pattern, we will also compute the so-called mutual information, and these results show that only neighboring unstable periodic points share linearly growing entanglement in the heating phase, see Sec.~\ref{Sec:MI}.

We emphasize that, due to the maximum in \eqref{heating_rate} over essentially $\sum_{p} 4N_{p}$ different functions of $(\tau_{1}, \tau_{2})$, both smooth and nonanalytic behaviors are possible.
Crucially, one can have cusps in $\nu$ also within the overall heating phase, see Fig.~\ref{Fig:SinCos_CFT_1}, and this possibility can also be demonstrated analytically, see Appendix~\ref{App:Analytical_results:Ex4}, but in general there can also be paths that avoid such cusps.

Finally, we note that heating phases, since they are given by periodic points, correspond precisely to so-called Arnold tongues, well known in the study of classical dynamical systems \cite{dMvS:1993}.


\section{Main tools}
\label{Sec:Main_tools}


Below we show how to use the tools in \cite{GLM:2018, Moo1:2019} to analytically study the Floquet drive given by \eqref{U_F} and \eqref{H_1_2}. 


\subsection{Diffeomorphism representations}
\label{Sec:Main_tools:Diffeomorphism representations}


First, we introduce
\begin{equation}
\label{f_j}
f_{j}(x)
= \int_{0}^{x} \dd x'\, \frac{v_{j, 0}}{v_{j}(x')},
\qquad
\frac{1}{v_{j, 0}}
= \frac{1}{L} \int_{-L/2}^{L/2} \frac{\dd x'}{v_{j}(x')}
\end{equation}
for $j = 1, 2$.
These are diffeomorphisms satisfying $f_{j}(x + L) = f_{j}(x) + L$ and $f_{j}'(x) > 0$.
More specifically, $f_{j}$ lie in the covering group $\wDiff_{+}(S^1)$ of orientation-preserving diffeomorphisms of the circle $S^1$ such that
\begin{multline}
\label{UHUm1}
U_{+}(f_{j}) U_{-}(f_{j}) H_{j} U_{-}(f_{j})^{-1} U_{+}(f_{j})^{-1} \\
= \int_{-L/2}^{L/2} \dd x\, v_{j, 0} \bigl[ T_{+}(x) + T_{-}(x) \bigr]
	+ \mathrm{const},
\end{multline}
using the projective unitary representations $U_{\pm}(f)$ of $f \in \wDiff_{+}(S^1)$ in \cite{GLM:2018, Moo1:2019}.
We note that their generators are precisely the energy-momentum components $T_{\pm}(x)$:
\begin{equation}
U_{\pm}(f)
= I \mp \ii \eps \int_{-L/2}^{L/2} \dd x\, \zeta(x) T_{\pm}(x) + o(\eps)
\end{equation}
for infinitesimal $f(x) = x + \eps \zeta(x)$.

Second, similar to \cite{Moo1:2019}, the diffeomorphisms in \eqref{f_j} can be used to compute exact analytical results for the Floquet time evolution of local observables
\begin{equation}
\cO(x; t)
= U_{F}^{-n} \cO(x) U_{F}^{n},
\qquad
t = n t_{\mathrm{cyc}},
\;
n \in \mathbb{Z}.
\end{equation}
It follows from \eqref{UHUm1} and known results for the adjoint action of $U_{\pm}(f_{j})$ \cite{Note5} that the time evolution is encoded into generalized light-cone coordinates $\tx_{t}^{\mp}(x)$ given by \cite{Note6}
\begin{equation}
\label{recursion}
\tx_{t + t_{\mathrm{cyc}}}^{\mp}(x)
= f_{\pm} ( \tx_{t}^{\mp}(x) ),
\qquad
\tx_{0}^{\mp}(x) = x,
\end{equation}
using
\begin{equation}
\label{f_pm}
f_{\pm}(x)
= f_{2}^{-1} \bigl(
		f_{2} \bigl(
			f_{1}^{-1}( f_{1}(x) \mp v_{1, 0}t_{1} )
		\bigr) \mp v_{2, 0}t_{2}
	\bigr).
\end{equation}
By definition, $f_{\pm} \in \wDiff_{+}(S^1)$, and this also implies that
$f_{\pm}^{p} \in \wDiff_{+}(S^1)$
for $f_{\pm}^{p}$ defined in \eqref{p-periodic_points}.

For instance, for a primary field $\Phi(x) = \Phi(x, x)$ with conformal weights $(\Delta^{+}, \Delta^{-})$ \cite{Note7},
\begin{equation}
\label{Phi_x_t}
\Phi(x; t)
= \biggl[ \frac{\partial \tx_{t}^{-}(x)}{\partial x} \biggr]^{\Delta^{+}}
	\biggl[ \frac{\partial \tx_{t}^{+}(x)}{\partial x} \biggr]^{\Delta^{-}}
	\!\!\!
	\Phi(\tx_{t}^{-}(x), \tx_{t}^{+}(x)),
\end{equation}
and for the components of the energy-momentum tensor,
\begin{equation}
\label{Tpm_x_t}
T_{\pm}(x; t)
= \biggl[ \frac{\partial \tx_{t}^{\mp}(x)}{\partial x} \biggr]^{2}
	T_{\pm}(\tx_{t}^{\mp}(x))
	- \frac{c}{24\pi} \bigl\{ \tx_{t}^{\mp}(x), x \bigr\},
\end{equation}
where
$\partial \tx_{t}^{\mp}(x) / \partial x
= \prod_{m = 0}^{n-1} f_{\pm}'(\tx_{m t_{\mathrm{cyc}}}^{\mp}(x))$ for $n > 0$ and
$\{ f(x), x \}$ denotes the Schwarzian derivative of $f(x)$ with respect to $x$.
Both results are important as they will allow us to compute correlation functions and the heating rate.


\subsection{Periodic points}

\begin{figure}[!htbp]

\centering

\includegraphics[scale=1, trim=0 0 0 0, clip=true]
{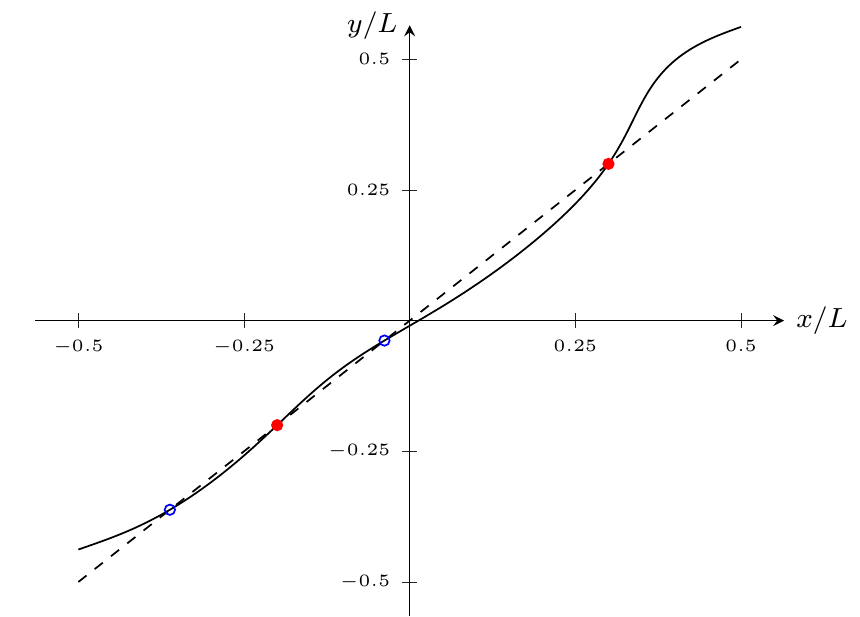}%

\vspace{-2.0mm}

\caption{%
Illustration of fixed points as intersections between the (black solid) curve $y = f_{+}(x)$ and the straight (black dotted) line $y = x$ (up to integer multiples of $L$).
Stable and unstable fixed points are indicated by empty blue and filled red circles, respectively.
[The plotted curve is for the case in Fig.~\ref{Fig:SinCos_CFT_1} at $(\tau_{1}, \tau_{2}) = (0.10, 0.45)$, which corresponds to a point in one of the darker leaves within the center leaf in Fig.~\ref{Fig:SinCos_CFT_1}(a), see also column (i) in Figs.~\ref{Fig:SinCos_CFT_2} and~\ref{Fig:SinCos_CFT_3}.]%
}

\label{Fig:Intersections}

\end{figure}

Periodic points are solutions $\tx_{\ast p}^{\mp}$ to \eqref{p-periodic_points} and correspond to intersections between the straight line $y = x$ and curves $y = f_{\pm}^{p}(x)$ (up to integer multiples of $L$ since we are on the circle).
This is illustrated in Fig.~\ref{Fig:Intersections} using the case in Fig.~\ref{Fig:SinCos_CFT_1} as an example.
Note that a periodic point of $f_{\pm}$ with period $p > 1$ can be viewed as a fixed point of $f_{\pm}^{p}$, and vice versa.

The following definition will be important:
Assuming $\tx_{\ast p}^{\mp}$ solves \eqref{p-periodic_points},
we say that it is
\begin{enumerate}[labelindent=1em, nosep, label={\textnormal{\arabic*.}}, ref={\textnormal{\arabic*}}]

\item
\emph{stable} if $f_{\pm}^{p \,\prime}(\tx_{\ast}^{\mp}) < 1$,

\item
\emph{unstable} if $f_{\pm}^{p \,\prime}(\tx_{\ast}^{\mp}) > 1$, or

\item
a \emph{tangent point} if $f_{\pm}^{p \,\prime}(\tx_{\ast}^{\mp}) = 1$.

\end{enumerate}
As a trivial example, note that $t_{1} = t_{2} = 0$, or equivalently $\tau_{1} = \tau_{2} = 0$, corresponds to a tangent point.

In Sec.~\ref{Sec:Main_tools:Properties_pps} we state a number of general properties for our periodic points.
In particular, apart from tangent points, they must come in pairs, one stable and one unstable.
Their existence dramatically affects the dynamics, see Fig.~\ref{Fig:SinCos_CFT_2} for typical examples.
To understand this, suppose that $x$ is a periodic point $\tx_{\ast p}^{\mp}$, then
$\partial \tx_{t}^{\mp}(x) / \partial x
= f_{\pm}^{p \,\prime}(\tx_{\ast p}^{\mp})^{n/p}$
for $t = n t_{\mathrm{cyc}}$ with $n$ a multiple of $p$.
Thus, if $f_{\pm}^{p \,\prime}(\tx_{\ast p}^{\mp}) > 1$ ($< 1$), this factor diverges (vanishes) exponentially as $t \to \io$.

At $(\tau_{1}, \tau_{2})$ where pairs of stable and unstable periodic points merge, the straight line and the $f_{\pm}^{p}$-curve are tangent at some $\tx_{\mathrm{c} p}^{\mp} = \tx_{\ast p}^{\mp}$ solving \eqref{p-periodic_points} and \eqref{tangent_points}.
This yields an equation system for $\tau_{1}$ and $\tau_{2}$ for each tangent point, cf.\ Appendix~\ref{App:Practical_eqs}.
Crucially, such tangent points correspond to curves separating regions in $(\tau_{1}, \tau_{2})$-space with different numbers of periodic points.
These regions (accounting for the symmetries illustrated in Fig.~\ref{Fig:Phase_diagrams}) typically have the shape of nested leaves and together give the overall heating phase, see, e.g., Figs.~\ref{Fig:Phase_diagrams} and~\ref{Fig:SinCos_CFT_1}.
Additional examples of heating phases due to fixed points only are given in Fig.~\ref{Fig:Phase_diagrams_fp_SinCos_CFT} in Appendix~\ref{App:Analytical_results}.


\subsection{Properties of periodic points}
\label{Sec:Main_tools:Properties_pps}


The solutions $\tx_{\ast p}^{\mp}$ to \eqref{p-periodic_points} have the following properties:
\begin{enumerate}[labelindent=1em, nosep, label={\textnormal{(\roman*)}}, ref={\textnormal{(\roman*)}}]

\item
\label{Item:US_pairs}
For each of the two $f_{\pm}$, stable and unstable periodic points come in pairs.
Moreover, up to tangent points (which can be viewed as a pair), periodic points must alternate between stable and unstable, i.e., there is always a stable point between two unstable ones and vice versa. 

\item
\label{Item:Convergence}
The convergence of $\tx_{t}^{\mp}(x)$ to a stable periodic point $\tx_{\ast p}^{\mp}$ (viewed as the evolution under $f_{\pm}^{p}$ covering $p$ integer time steps) is exponentially fast, with the rate given by $\log[f_{\pm}^{p \,\prime}(\tx_{\ast p}^{\mp})]$.

\item
\label{Item:Time_rev}
For each solution of \eqref{p-periodic_points} for given $(t_{1}, t_{2})$, there is also a solution for $(-t_{1}, -t_{2})$.

\item
\label{Item:Shifts}
Shifts of $t_{1}$ by $L/k_{2}v_{1, 0}$ and $t_{2}$ by $L/k_{1}v_{2, 0}$ leave \eqref{p-periodic_points} invariant, where $L/k_{j}$ for $k_{j} \in \mathbb{Z}^{+}$ are the periods of $v_{j}(x)$; we set $k_{j} = 1$ if $v_{j}(x)$ is constant.

\end{enumerate}
Proofs are given in Appendix~\ref{App:Proof_of_properties_pps}.

We stress that Properties~\ref{Item:Time_rev} and \ref{Item:Shifts} imply the symmetries illustrated in Fig.~\ref{Fig:Phase_diagrams} by considering $\tau_{1, 2} = v_{1, 2} t_{1, 2}/L$ instead of $t_{1, 2}$.

For regions in $(\tau_{1}, \tau_{2})$-space with periodic points with period $p$, it is sometimes natural to consider the evolution in units of $p$ integer time steps.
Note that, on general grounds, since we work with smooth deformations defining circle diffeomorphisms, a given region in the phase diagram cannot have periodic points with different periods \cite{dMvS:1993}.


\section{Energy density and flow of excitations}
\label{Sec:cE_and_2pcf}


The time-evolved energy density corresponding to $H_{1}$ in \eqref{H_1_2} is
\begin{equation}
\label{cE_1}
\cE_{1}(x; t)
= v_{1}(x) \bigl[ T_{+}(x; t) + T_{-}(x; t) \bigr].
\end{equation}
For simplicity, we will use this quantity to study the flow of energy in our Floquet system.
Note that we may as well use $\cE_{2}(x; t)$ with $v_{1}(x)$ replaced by $v_{2}(x)$; indeed, for our purposes, the exact choice is less important as long as it is a linear combination of both $T_{\pm}(x; t)$.

\begin{figure}[!htbp]

\centering

\includegraphics[scale=1, trim=0 0 0 0, clip=true]
{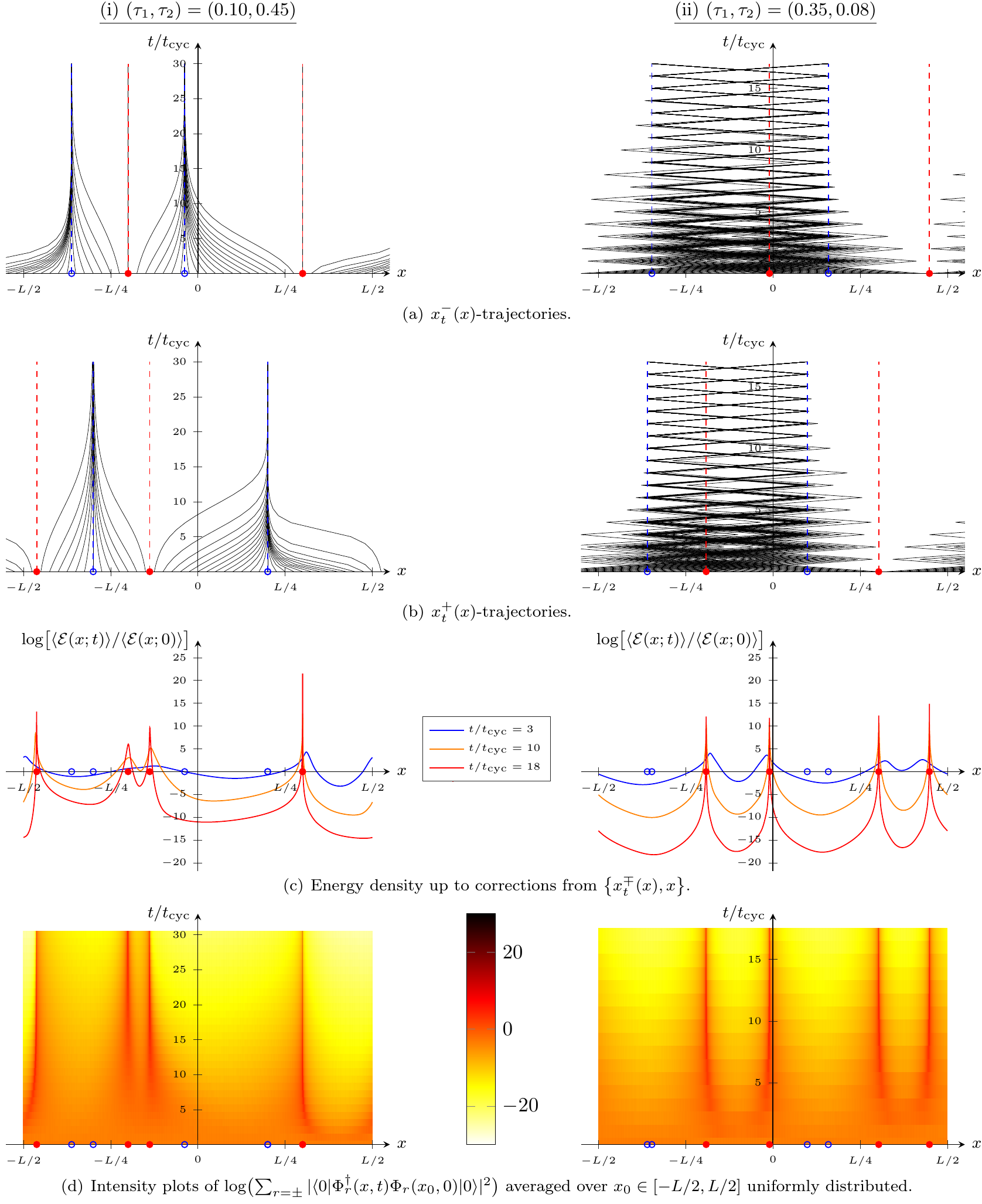}%

\vspace{-2.0mm}

\caption{%
Plots for the case in Fig.~\ref{Fig:SinCos_CFT_1} at
(i) $(\tau_{1}, \tau_{2}) = (0.10, 0.45)$ in the left column
and
(ii) $(\tau_{1}, \tau_{2}) = (0.35, 0.08)$ in the right column.
In (i) and (ii) there are eight fixed points and eight periodic points with period 2, respectively.
(a,~b)
$\tx_{t}^{\mp}(x)$-trajectories.
(c)
Energy-density expectation $\langle \cE(x; t) \rangle$ for
$\cE(x; t)
= \cE_{1}(x; t) + (c v_{1} / 24 \pi) [ \{ \tx_{t}^{-}(x), x \} + \{ \tx_{t}^{+}(x), x \} ]$ given by \eqref{cE_1} with respect to an arbitrary translation-invariant state.
(d)
Correlation functions $\langle0| \Phi^{\dagger}_{r}(x, t) \Phi_{r}(x_{0}, 0) |0\rangle$ with respect to the ground state $|0\rangle$ of $H_{1}$ for primary fields $\Phi_{r}(x, t)$ with $\Delta_{r}^{\pm} = \delta_{r, \pm}$ for $r = \pm$, averaged over $x_{0} \in [-L/2, L/2]$ uniformly distributed, setting $L = 50$.
Stable and unstable periodic points are indicated by empty blue and filled red circles, respectively.%
}

\label{Fig:SinCos_CFT_2}

\end{figure}

The properties of our periodic points and \eqref{cE_1} together with \eqref{Tpm_x_t} imply that the energy density $\cE_{1}(x; t)$ grows (decays) exponentially in time at unstable (stable) such points.
Moreover, if there are periodic points, all $x$ except the unstable points flow exponentially fast to the stable ones, see Figs.~\ref{Fig:SinCos_CFT_2}(a) and~\ref{Fig:SinCos_CFT_2}(b) for typical examples of this behavior.
This follows from the general properties of our periodic points in Sec.~\ref{Sec:Main_tools:Properties_pps}.
As a consequence, the whole system is cooling apart from the exceptions which are heating exponentially, see Fig.~\ref{Fig:SinCos_CFT_2}(c) where we plot $\cE_{1}(x; t)$ in \eqref{cE_1} up to the Schwarzian-derivative terms in \eqref{Tpm_x_t}; these terms do not affect the existence or location of periodic points, cf.\ \cite{LLMM2:2017, GLM:2018, Moo1:2019} for discussions of similar Schwarzian-derivative contributions.
It follows by integrating $\cE_{1}(x, t)$ over $x \in [-L/2, L/2]$ that the presence of unstable points leads to an overall heating of the system with the heating rate $\nu$ determined by the unstable point with the steepest slope, which implies \eqref{heating_rate}.
On the other hand, the absence of periodic points implies $\nu = 0$ by definition, which characterizes the nonheating phase, and pseudoperiodic evolution of all trajectories \cite{Note8}.

Lastly, by computing two-point correlation functions for primary fields one can study the flow of excitations in the system.
These follow from \eqref{Phi_x_t} and known results for homogeneous CFT:
To compute correlation functions for any primary field, all that is needed is knowledge about the corresponding correlation function for homogeneous CFT, cf.\ \cite{GLM:2018, Moo1:2019}.
For simplicity, consider the ground state $|0\rangle$ of the homogeneous theory.
Then, under our Floquet time evolution, the results for two-point correla-

\onecolumngrid

\pagebreak

\twocolumngrid

\noindent
tion functions with respect to $|0\rangle$ are
\begin{multline}
\label{Phi_2pcf}
\langle0| \Phi(x; t)^{\dagger} \Phi(x'; t') |0\rangle \\
\begin{aligned}
& = \biggl[
			\frac{\partial \tx_{t}^{-}(x)}{\partial x }
			\frac{\partial \tx_{t'}^{-}(x')}{\partial x'}
		\biggr]^{\Delta^{+}}
		\biggl[
			\frac{\partial \tx_{t}^{+}(x)}{\partial x }
			\frac{\partial \tx_{t'}^{+}(x')}{\partial x'}
		\biggr]^{\Delta^{-}} \\
& \quad \times
		\biggl(
			\frac{ \ii \pi}{L \sin( \pi [\tx_{t}^{-}(x) - \tx_{t'}^{-}(x') + \ii 0^+]/L )}
		\biggr)^{2\Delta^{+}} \\
& \quad \times
		\biggl(
			\frac{-\ii \pi}{L \sin( \pi [\tx_{t}^{+}(x) - \tx_{t'}^{+}(x') - \ii 0^+]/L )}
		\biggr)^{2\Delta^{-}}
\end{aligned}
\end{multline}
for finite $L$.
This result together with the properties in Sec.~\ref{Sec:Main_tools:Properties_pps} shows that, while almost all $\tx_{t}^{\mp}(x)$-trajectories flow to stable periodic points in the heating phase, the probability density of excitations flow exponentially fast to the unstable ones, see Fig.~\ref{Fig:SinCos_CFT_2}(d) for examples of this.


\section{Entanglement entropy}
\label{Sec:EE}


As is commonplace, the von Neumann entanglement entropy $S_{A}(t)$ for the subsystem on the interval $A = [x, y]$ with the rest can be computed from correlation functions for twist fields \cite{CCaD:2008, CaCa2:2016}.
We show in Appendix~\ref{App:EE} that
\begin{align}
S_{A}(t)
& = \frac{c}{12} \bigl[ S_{+}(t) + S_{-}(t) \bigr],
		\label{EE} \\
S_{\pm}(t)
& = - \log
			\Biggl[ \!
				\frac{\partial \tx_{t}^{\mp}}{\partial x}
				\frac{\partial \ty_{t}^{\mp}}{\partial y}
				\biggl(
					\frac{ \pm \ii \pi}
						{L \sin \bigl( \frac{\pi}{L} [\tx_{t}^{\mp} - \ty_{t}^{\mp} \pm \ii 0^+] \bigr)}
				\biggr)^{2}
			\Biggr] \nonumber
\end{align}
with $\tx_{t}^{\mp} = \tx_{t}^{\mp}(x)$ and $\ty_{t}^{\mp} = \tx_{t}^{\mp}(y)$.
Its behavior is different depending on the absence or presence of periodic points.
In the nonheating phase, as mentioned, there is an underlying pseudoperiodicity in time.
On the other hand, when heating, the situation is drastically different.

\begin{figure}[!htbp]

\centering

\includegraphics[scale=1, trim=0 0 0 0, clip=true]
{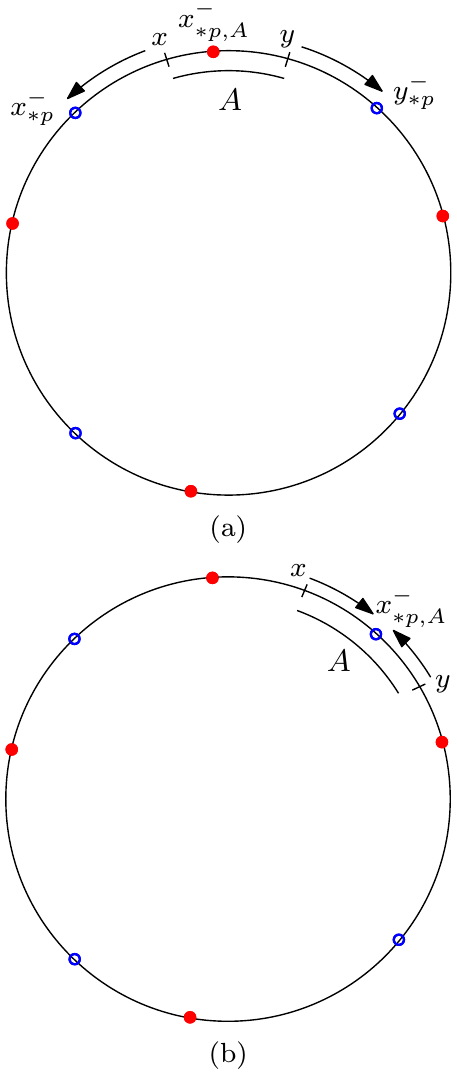}%

\vspace{-2.0mm}

\caption{%
Illustrations of the two cases (a) and (b) described in the text for the entanglement entropy $S_{A}(t)$ in the heating phase.
Stable and unstable periodic points are indicated by empty blue and filled red circles, respectively.
For simplicity, we illustrate only the contribution from the right-moving component.
(Flows to periodic points with period $p$ should be interpreted as those under $f_{+}^{p}$.)%
}

\label{Fig:Entanglement_pattern_Case_ab}

\end{figure}

Using the properties of our periodic points in Sec.~\ref{Sec:Main_tools:Properties_pps}, we can study the behavior of $S_{A}(t)$ in the heating phase.
We consider the following two cases, illustrated in Fig.~\ref{Fig:Entanglement_pattern_Case_ab}:
$A$ contains
(a)
at least one of (but not all) the unstable periodic points
or
(b)
no (or all) unstable periodic points.
(Containing ``no'' or ``all'' points are equivalent due to our periodic boundary conditions.)
For simplicity, and without loss of generality since the two $\pm$-components commute, we consider only the right-moving ($+$) component:

(a)
Due to Property~\ref{Item:US_pairs} in Sec.~\ref{Sec:Main_tools:Properties_pps},
$\tx_{t}^{-}$ and $\ty_{t}^{-}$ flow to two different stable periodic points,
denoted $\tx_{\ast p}^{-}$ and $\ty_{\ast p}^{-}$, respectively.
(For simplicity, we consider the flow in units of $p$ integer time steps.)
Since $\tx_{\ast p}^{-} \neq \ty_{\ast p}^{-}$, the difference $\tx_{t}^{-} - \ty_{t}^{-}$ in \eqref{EE} remains finite as $t \to \infty$, while the derivative factors decay exponentially to $0$ as $\tx_{t}^{-}$ and $\ty_{t}^{-}$ flow to their respective stable periodic points.
It follows that the leading contribution to $S_{+}(t)$ is
\begin{equation}
\label{cS_EE_1upp_lt}
- \frac{
		\log \bigl[ f_{+}^{p \,\prime}(\tx_{\ast p}^{-}) \bigr]
		+ \log \bigl[ f_{+}^{p \,\prime}(\ty_{\ast p}^{-}) \bigr]
	}{p t_{\mathrm{cyc}}} \, t,
\end{equation}
i.e., $S_{+}(t)$ grows linearly in time for large $t$.

(b)
Due to the same property as in (a), $\tx_{t}^{-}$ and $\ty_{t}^{-}$ flow to the same stable periodic point, denoted $\tx_{\ast p, A}^{-}$.
It follows from this and Property~\ref{Item:Convergence} in Sec.~\ref{Sec:Main_tools:Properties_pps} that $\tx_{t}^{-} - \ty_{t}^{-}$ in \eqref{EE} decays exponentially at a rate that exactly cancels the exponential decay of the derivative factors, implying that $S_{+}(t)$ becomes constant for large $t$.

Note that, for all cases, if $t = 0$, then $\tx_{0}^{\mp}(x) = x$, which means that \eqref{EE} yields
\begin{equation}
S_{A}(0)
= ({c}/{3})
	\log
	\bigl[
		({L}/{\pi}) \sin \bigl( {\pi \ell}/{L} \bigr)
	\bigr]
\end{equation}
for $\ell = x - y > 0$.
This is precisely the usual equilibrium result in the case of periodic boundary conditions.

\begin{figure}[!htbp]

\centering

\vspace{-1.0mm}

\includegraphics[scale=1, trim=0 0 0 0, clip=true]
{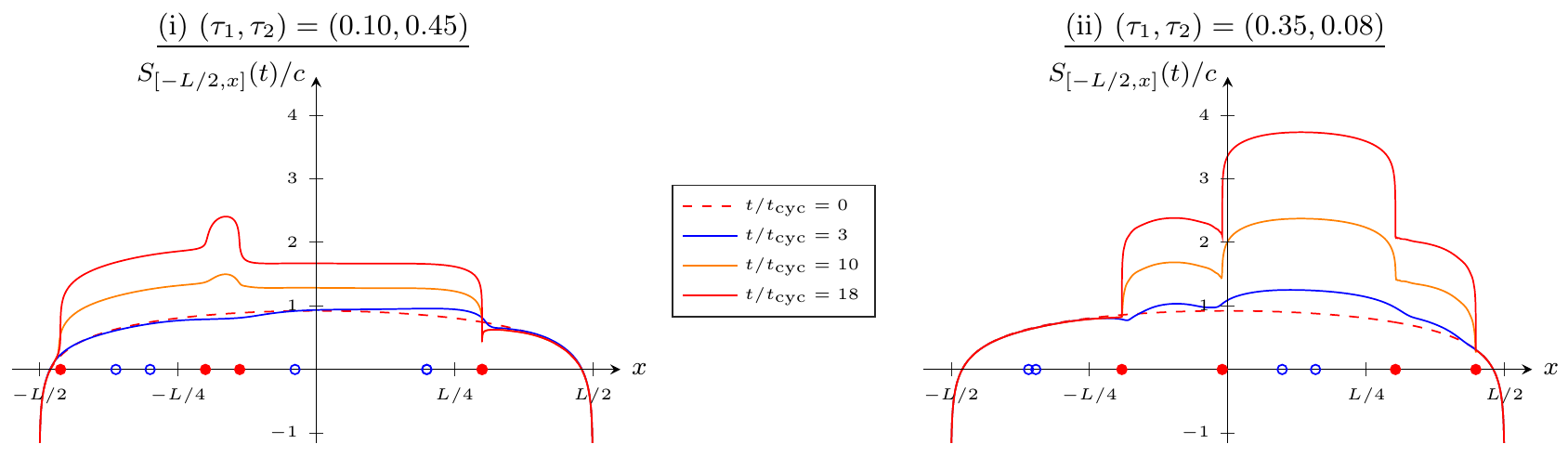}%

\vspace{-2.0mm}

\caption{%
Plots of the entanglement entropy $S_{[-L/2, x]}(t)$ given by \eqref{EE}, setting $L = 50$, for the case in Fig.~\ref{Fig:SinCos_CFT_1} at
(i) $(\tau_{1}, \tau_{2}) = (0.10, 0.45)$
and
(ii) $(\tau_{1}, \tau_{2}) = (0.35, 0.08)$.
See the caption to Fig.~\ref{Fig:SinCos_CFT_2} for further details.%
}%

\label{Fig:SinCos_CFT_3}

\end{figure}

The exact analytical results in \eqref{EE} give rise to a remarkable structure of kinks in the entanglement entropy $S_{[-L/2,x]}(t)$ that brings to mind Lifshitz phase transitions \cite{RSLLS:2013}, see Fig.~\ref{Fig:SinCos_CFT_3} for examples of such structures.
Indeed, when crossing a transition line between regions with different numbers of unstable periodic points, cf.\ Fig.~\ref{Fig:SinCos_CFT_1}(a), the number of kinks in $S_{[-L/2,x]}(t)$ changes by an even integer.
This can be viewed as a nonequilibrium analogue of a Lifshitz phase transition:
Instead of jumps or cusps

\onecolumngrid

\twocolumngrid

\noindent
\cite{RSLLS:2013} there are, in general, kinks in the entanglement entropy induced by the unstable periodic points.
Note that such transitions within the heating phase are local and do not affect global properties such as the heating rate.


\section{Mutual information}
\label{Sec:MI}


To understand the entanglement-entropy pattern, we study the mutual information
$I_{A; B}(t) = S_{A}(t) + S_{B}(t) - S_{A \cup B}(t)$
for two intervals $A = [x_1, x_2]$ and $B = [x_3, x_4]$.
One can show that $I_{A; B}(t)$ evolves nontrivially in time if and only if $A$ and $B$ contain at least one unstable periodic point each, separated by a single stable point (with respect to the same component).
As in Sec.~\ref{Sec:EE}, for simplicity, and without loss of generality, we consider only the contributions from the right-moving component.

\begin{figure}[!htbp]

\centering

\includegraphics[scale=1, trim=0 0 0 0, clip=true]
{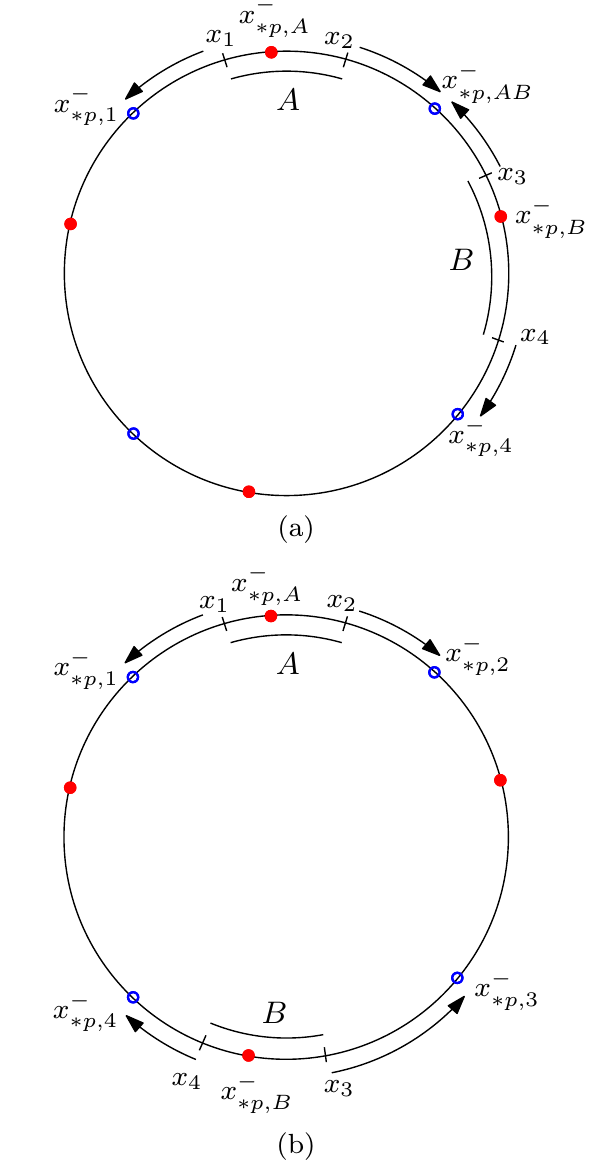}%

\vspace{-2.0mm}

\caption{%
Illustrations of the two cases (a) and (b) described in the text for the mutual information $I_{A; B}(t)$ in the heating phase.
See the caption to Fig.~\ref{Fig:Entanglement_pattern_Case_ab} for further details.%
}

\label{Fig:Mutual_info}

\vspace{-3.0mm}

\end{figure}

More specifically, in Appendix~\ref{App:MI}, we compute $I_{A; B}(t)$ for the following two cases, illustrated in Fig.~\ref{Fig:Mutual_info}:

(a)
Suppose that $A$ and $B$ are neighbors, i.e., that there is only one stable periodic point, denoted $\tx_{\ast p, AB}^{-}$, between them, see Fig.~\ref{Fig:Mutual_info}(a).
It follows that $\tx_{t}^{-}(x_2)$ and $\tx_{t}^{-}(x_3)$ flow to $\tx_{\ast p, AB}^{-}$, while $\tx_{t}^{-}(x_1)$ and $\tx_{t}^{-}(x_4)$ flow to two other different stable periodic points, denoted $\tx_{\ast p, 1}^{-}$ and $\tx_{\ast p, 4}^{-}$, respectively.
In this case, we show that the leading contribution to $I_{A; B}(t)$ from the right-moving component is
\begin{equation}
\label{MI_rmc_case_a}
- \frac{c}{6} \frac{\log[f_{+}^{p \,\prime}(\tx_{\ast p, AB}^{-})]}{p t_{\mathrm{cyc}}} \, t,
\end{equation}
i.e., it grows linearly in time with a rate given by the stable periodic point between $A$ and $B$.

(b)
Suppose that $A$ and $B$ are not neighbors, i.e., there is not only a single stable periodic point separating them, see Fig.~\ref{Fig:Mutual_info}(b).
(Such a situation can only occur if there are at least four unstable periodic points for each component.)
In this case, all $\tx_{t}^{-}(x_{j})$ ($j = 1, 2, 3, 4$) flow to four different stable periodic points.
We show that this leads to a cancellation between $S_{A \cup B}(t)$ and $S_{A}(t) + S_{B}(t)$ such that the contribution from the right-moving component to $I_{A; B}(t)$ does not change with time for large $t$.

As a special case, suppose that $A$ and $B$ contain exactly one unstable point each.
Our results then imply that only neighboring unstable periodic points share entanglement that grows linearly at late times.
This generalizes previous results obtained in \cite{FGVW:2020} for the special case of $\mathfrak{sl}(2)$-deformations to inhomogeneous CFT with general smooth deformations.
Moreover, if the number of such points for each component is even \cite{Note9}, then the entanglement entropy is ``bipartite'' \cite{Note10}.


\section{Concluding remarks}
\label{Sec:Concluding_remarks}


We studied two-step Floquet drives in 1+1-dimen\-sional many-body systems described by inhomogeneous CFT, generalizing previous results for a subfamily of similar systems to general smooth deformations.
To do this, we developed a new geometric approach based on an exact correspondence with dynamical systems on the circle:
We showed that one can encode the time evolution into coordinate transformations given by orientation-preserving diffeomorphisms of the circle obtained directly from the deformations.
In addition, we emphasized that our drive is a simple example and that our approach can be straightforwardly applied to, e.g., multistep, random, chaotic, and quasiperiodic drives.
We then showed that the presence of fixed or higher-periodic points of these transformations implies that the system is heating, with (nested) leaf-shaped heating phases characterized by unstable such points, while their absence means that the system is in a nonheating phase.
The heating rate was identified as the natural order parameter and shown to be determined by the ``most unstable'' periodic point, implying that it can change nonanalytically even within the overall heating phase.
As an example of a local property we computed two-point correlation functions and used this to show that excitations (as well as energy) concentrate exponentially fast at unstable periodic points.
Lastly, we studied the entanglement entropy and the mutual information.
We showed that unstable periodic points lead to a remarkable structure of kinks in the entanglement entropy, reminiscent of Lifshitz phase transitions,
and that only neighboring unstable periodic points share linearly growing entanglement.

We remark that our approach is suitable for numerical implementations (cf.\ Fig.~\ref{Fig:SinCos_CFT_1}, where parts of the phase diagram were obtained numerically).
As argued, we also expect that our approach and underlying ideas are useful practically and as inspiration even beyond CFT.
It would be interesting to explore this further and to compare with exact numerical results, e.g., for inhomogeneous $XXZ$ spin chains.
Additionally, it would be interesting to study effects of different boundary conditions.

Essential inspiration to our geometric approach comes from minimal surfaces:
Similar equations appear when fitting such a surface to given boundary conditions \cite{Note11} and the existence of fixed-point solutions corresponds to when it extremizes the surface area.
It would be interesting to explore if there are deeper such connections with variational problems.
Moreover, as our geometric approach is applicable to any 1+1-dimensional CFT, it should also be applicable to large-central-charge ones.
These classes of CFTs are often studied using holography, as predicted by the AdS/CFT correspondence.
It would therefore be desirable to obtain the AdS dual to such CFTs in our Floquet set-up and to understand the emergence of heating and nonheating phases from a gravitational point of view.
Finally, it would be interesting to use, e.g., some of the many fixed-point theorems in the mathematics literature to develop a potential complete characterization of phase diagrams.


\begin{acknowledgments}
We thank Ramasubramanian Chitra, Kenny Choo, Nicol{\`o} Defenu, Krzysztof Gaw\k{e}dzki, Gian Michele Graf, Titus Neupert, Cl{\'e}ment Tauber, and Apoorv Tiwari for fruitful discussions.
P.M.\ is particularly thankful to Jens Hoppe for teaching him about minimal surfaces, which became an invaluable inspiration to the approach in this paper.
P.M.\ is grateful for financial support from the Wenner-Gren Foundations (Grant No.\ WGF2019-0061).
B.L.\ gratefully acknowledges financial support from the European Research Council (ERC) under the European Union's Horizon 2020 research and innovation program (ERC Starting Grant PARATOP, Grant Agreement No.\ 757867).
\end{acknowledgments}


\begin{appendix}


\section{Proof of properties for periodic points}
\label{App:Proof_of_properties_pps}


In what follows, we give brief proofs of the properties of our periodic points listed in Sec.~\ref{Sec:Main_tools:Properties_pps}.
For simplicity, we prove them for fixed points $\tx_{\ast}^{\mp} = \tx_{\ast 1}^{\mp}$;
the properties for periodic points with period $p = 2, 3, \ldots$ follow by replacing $f_{\pm}$ in \eqref{f_pm} by $f_{\pm}^{p}$ in \eqref{p-periodic_points}.

To prove \ref{Item:US_pairs}, we note that $f_{\pm}(-L/2) \lessgtr -L/2$ is equivalent to $f_{\pm}(L/2) \lessgtr L/2$ since $f_{\pm} \in \wDiff_{+}(S^1)$, and thus, if the curve $y = f_{\pm}(x)$ crosses the straight line $y = x$ at a point where $f_{\pm}' \lessgtr 1$, it must cross this line again at another point where $f_{\pm}' \gtrless 1$.
Continuity of $f_{\pm}$ ensures the alternating appearance of the fixed points.
\ref{Item:Convergence} is a standard property of stable fixed points; to find the rate, one can use the mean value theorem to conclude that
$\lim_{m\to\infty} |\tx_{(m+1)t_{\mathrm{cyc}}}^{\mp}(x) - \tx_{\ast}^{\mp}| / |\tx_{mt_{\mathrm{cyc}}}^{\mp}(x) - \tx_{\ast}^{\mp}| = f_{\pm}'(\tx_{\ast}^{\mp}) < 1$.
To prove \ref{Item:Time_rev}, note that if $\tx_{\ast}^{\mp} = f_{\pm}(\tx_{\ast}^{\mp})$ then also $f_{\mp}(\ty_{\ast}^{\pm}) = \ty_{\ast}^{\pm}$ by setting $\tx_{\ast}^{\mp} = f_{1}^{-1}( f_{1}(\ty_{\ast}^{\pm}) \pm v_{1, 0}t_{1} )$ and rearranging.
\ref{Item:Shifts} follows using \eqref{f_j}, \eqref{f_pm}, and $f_{j} \in \wDiff_{+}(S^1)$.


\section{Practical equations for fixed points}
\label{App:Practical_eqs}


For $p = 1$, the formulas in \eqref{p-periodic_points} and \eqref{tangent_points} can be written in a symmetric way for both $\pm$-components as follows.

First, define
\begin{equation}
\label{F_j}
F_{j}(\xi) = \int_{0}^{\xi} \dd \xi'\, \frac{1}{w_{j}(\xi')}
\end{equation}
for $j = 1, 2$.
It follows that $F_{j}(\xi) = (v_{j}/Lv_{j, 0}) f_{j}(\xi L)$.

Second, define $\xi_{\ast}$ by
$L (v_{1, 0}/v_{1}) \xi_{\ast} = f_{1}(\tx_{\ast}^{\mp}) \mp v_{1, 0}t_{1}/2$.
Using this, any fixed points $\tx_{\ast}^{\mp} = \tx_{\ast 1}^{\mp}$ can be expressed as
\begin{equation}
\label{x_mp_ast_xi_ast}
\tx_{\ast}^{\mp}
= f_{1}^{-1}( L v_{1, 0} \xi_{\ast}/v_{1} \pm v_{1, 0} t_{1}/2 )
= L F_{1}^{-1}( \xi_{\ast} \pm \tau_{1}/2 ).
\end{equation}

Third, inserting \eqref{x_mp_ast_xi_ast} into \eqref{p-periodic_points} and rearranging implies that $\tx_{\ast}^{\mp}$ are given by solutions $\xi_{\ast}$ to
\begin{equation}
\label{tau_2_xi_ast_general}
\tau_{2}
= \int^{\xi_{\ast} - \tau_{1}/2}_{\xi_{\ast} + \tau_{1}/2}
	\dd \xi'\, \frac{w_{1}(F_{1}^{-1}(\xi'))}{w_{2}(F_{1}^{-1}(\xi'))}
\end{equation}
for given $(\tau_{1}, \tau_{2})$.

Fourth, from \eqref{f_pm} and the above, it follows that \eqref{tangent_points} is equivalent to that tangent points $\tx_{\mathrm{c}}^{\mp} = \tx_{\mathrm{c} 1}^{\mp}$ are given by \eqref{x_mp_ast_xi_ast} with $\xi_{\ast} = \xi_{\mathrm{c}}$ that also solves
\begin{equation}
\label{tau_1_xi_c_general}
\frac
	{w_{2}( F_{1}^{-1}( \xi_{\mathrm{c}} - \tau_{1}/2 ) )}
	{w_{1}( F_{1}^{-1}( \xi_{\mathrm{c}} - \tau_{1}/2 ) )}
= \frac
		{w_{2}( F_{1}^{-1}( \xi_{\mathrm{c}} + \tau_{1}/2 ) )}
		{w_{1}( F_{1}^{-1}( \xi_{\mathrm{c}} + \tau_{1}/2 ) )}.
\end{equation}

Combining \eqref{tau_2_xi_ast_general} and \eqref{tau_1_xi_c_general} gives an equation system for $(\tau_{1}, \tau_{2})$ for each solution $\xi_{\mathrm{c}}$.

For the special case when $v_{1}(x) = v_{1}$ is constant, \eqref{p-periodic_points} and \eqref{tangent_points} for $p = 1$ can be written as
\begin{equation}
\label{tau_2_xi_ast_special}
\tau_{2}
= \int^{\xi_{\ast} - \tau_{1}/2}_{\xi_{\ast} + \tau_{1}/2}
	\dd \xi'\, \frac{1}{w_{2}(\xi')}
= F_{2}(\xi_{\ast} - \tau_{1}/2) - F_{2}(\xi_{\ast} + \tau_{1}/2)
\end{equation}
and
\begin{equation}
\label{tau_1_xi_c_special}
w_{2}(\xi_{\mathrm{c}} - \tau_{1}/2)
= w_{2}(\xi_{\mathrm{c}} + \tau_{1}/2),
\end{equation}
respectively.

In general, it follows from the above that any fixed points for the right- and left-moving components (in addition to appearing in pairs of stable and unstable points within a component) come in pairs $\tx_{\ast}^{\mp}$ given by \eqref{x_mp_ast_xi_ast} split between the components.
This together with \eqref{f_pm} implies
\begin{equation}
\label{f_deriv_pm_x_mp_ast}
f_{\pm}'(\tx_{\ast}^{\mp})
= \frac{v_{2}(\tx_{\ast}^{\mp})}{v_{2}(\tx_{\ast}^{\pm})}
	\frac{v_{1}(\tx_{\ast}^{\pm})}{v_{1}(\tx_{\ast}^{\mp})},
\end{equation}
and thus $f_{+}'(\tx_{\ast}^{-}) f_{-}'(\tx_{\ast}^{+}) = 1$ for each such pair.
In other words, if one of these fixed points is stable, then the other is unstable, and vice versa.

The above equations are useful for practical computations, but we emphasize that one must take into account the symmetries for $\tau_{1}$ and $\tau_{2}$ when working with them, see Sec.~\ref{Sec:Introduction} and Properties~\ref{Item:Time_rev} and~\ref{Item:Shifts} in Sec.~\ref{Sec:Main_tools:Properties_pps}.


\section{Analytical results for special cases}
\label{App:Analytical_results}


Below we study in greater detail two special cases of our two-step Floquet drive with $H_{1}$ homogeneous and $H_{2}$ given first by Ex.~1 and second by Ex.~4 in Table~\ref{Table:Examples}.


\subsection{Phase-transition lines for Ex.~1 in Table~\ref{Table:Examples}}
\label{App:Analytical_results:Ex1}


Consider $H_{2}$ given by the deformation of Ex.~1 in Table~\ref{Table:Examples}, i.e.,
\begin{equation}
w_{2}(\xi) = 1 + g \bigl[ 2\cos^2(\pi \xi) - 1 \bigr],
\end{equation}
which can be seen as a regularization of SSD CFT.
From \eqref{F_j} and $w_{j, 0} = v_{j, 0}/v_{j}$, we obtain
$F_{2}(\xi) = F_{g\mathrm{SSD}}(\xi)$ and $w_{2, 0} = w_{g\mathrm{SSD}, 0}$ with
\begin{subequations}
\label{F_and_w0_gSSD_CFT}
\begin{align}
F_{g\mathrm{SSD}}(\xi)
& = \frac{
			\arctan \bigl( \sqrt{(1 - g)/(1 + g)} \tan(\pi \xi) \bigr)
		}{
			\pi\sqrt{1 - g^2}
		}, \\
w_{g\mathrm{SSD}, 0}
& = \sqrt{1 - g^2}.
\end{align}
\end{subequations}
It is straightforward to check that
\begin{subequations}
\label{F_and_w0_SSD_CFT}
\begin{align}
F_{g\mathrm{SSD}}(\xi)
& \xrightarrow{g \to 1^{-}}
	F_{\mathrm{SSD}}(\xi)
	= \frac{\tan(\pi \xi)}{2\pi}, \\
w_{g\mathrm{SSD}, 0}
& \xrightarrow{g \to 1^{-}}
	w_{\mathrm{SSD}, 0}
	= 0,
\end{align}
\end{subequations}
and
\begin{equation}
\label{F_and_w0_sCFT}
F_{g\mathrm{SSD}}(\xi)
\xrightarrow{g \to 0^{+}}
\xi,
\qquad
w_{g\mathrm{SSD}, 0}
\xrightarrow{g \to 0^{+}}
1.
\end{equation}
The limiting case in \eqref{F_and_w0_SSD_CFT} explains why a regularization of SSD CFT is necessary:
For instance, $v_{2, 0} = v_{2} w_{g\mathrm{SSD}, 0}$ vanishes as $g \to 1^{-}$,
meaning that $f_{2}$ in \eqref{f_j} would not be well defined and thus not be an element of $\wDiff_{+}(S^1)$, cf.\ Sec.~\ref{Sec:Main_tools}.

For $g > 0$, the only nontrivial solutions to \eqref{tau_1_xi_c_special} are $\xi_{\mathrm{c}} = -1/2$ and $0$.
Inserting this together with \eqref{F_and_w0_gSSD_CFT} into \eqref{tau_2_xi_ast_special} and making our translation symmetries in $(\tau_{1}, \tau_{2})$-space manifest yields
\begin{equation}
\label{PT_lines_gSSD_CFT}
\tau_{2}
= \frac{
		n
	}{
		\sqrt{1 - g^2}
	}
	+
	\frac{
		2 \arctan \Bigl( \sqrt{\frac{1 - g}{1 + g}} \tan(\pi [m - \tau_{1}]/2) \Bigr)
	}{
		\pi\sqrt{1 - g^2}
	}
\end{equation}
for $m, n \in \mathbb{Z}$.
Clearly, it suffices to look at $(\tau_{1}, \tau_{2}) \in (0, 1) \times (0, 1/w_{g\mathrm{SSD}, 0})$, and in this ``unit cell'' the transition lines are given by
\begin{subequations}
\label{PT_lines_gSSD_CFT_curves12}
\begin{align}
\tau_{2}
& = \frac{
			2 \arctan \Bigl( \sqrt{\frac{1 - g}{1 + g}} \tan(\pi [1 - \tau_{1}]/2) \Bigr)
		}{
			\pi\sqrt{1 - g^2}
		},
		\label{PT_lines_gSSD_CFT_curve1} \\
\tau_{2}
& = \frac{1}{\sqrt{1 - g^2}}
		-
		\frac{
			2 \arctan \Bigl( \sqrt{\frac{1 - g}{1 + g}} \tan(\pi \tau_{1}/2) \Bigr)
		}{
			\pi\sqrt{1 - g^2}
		}.
		\label{PT_lines_gSSD_CFT_curve2}
\end{align}
\end{subequations}
In the limit $g \to 1^{-}$, similar to \eqref{F_and_w0_SSD_CFT}, it follows from \eqref{PT_lines_gSSD_CFT_curve1} that
\begin{equation}
\label{PT_lines_SSD_CFT}
\tau_{2}
= \frac{\tan(\pi [1 - \tau_{1}]/2)}{\pi},
\end{equation}
while \eqref{PT_lines_gSSD_CFT_curve2} disappears to infinity [except for $\tau_{1} \in \mathbb{Z}$ that give rise to straight vertical lines, see Fig.~\ref{Fig:Phase_diagrams}(b)].
In the other limiting case, $g \to 0^{+}$, similar to \eqref{F_and_w0_sCFT}, both \eqref{PT_lines_gSSD_CFT_curve1} and \eqref{PT_lines_gSSD_CFT_curve2} become
\begin{equation}
\label{PT_lines_sCFT}
\tau_{2} = 1 - \tau_{1},
\end{equation}
i.e., the phase diagram consists of a straight ``double'' line describing an ``empty'' transition.

The results in \eqref{PT_lines_gSSD_CFT_curves12} and \eqref{PT_lines_SSD_CFT} are plotted in Figs.~\ref{Fig:Phase_diagrams}(a) and~\ref{Fig:Phase_diagrams}(b) (invoking the symmetries in $\tau_{1}$ and $\tau_{2}$).
As stressed in the main text, Fig.~\ref{Fig:Phase_diagrams}(b) agrees perfectly with Fig.~1(c) in \cite{LCTTNC1:2020}, and one can even analytically show that \eqref{PT_lines_SSD_CFT} gives the same results as in the cited reference (using that the transition lines for SSD CFT in \cite{LCTTNC1:2020} are implicitly given by
$[1 - (\pi \tau_{2})^2] \sin^2(\pi \tau_{1}) + \pi \tau_{2} \sin(2\pi\tau_{1}) = 0$ 
with $\tau_{1} = T_{0}/L$ and $\tau_{2} = T_{1}/L$, translating the notation in \cite{LCTTNC1:2020} to ours).


\subsection{Phase diagrams due to fixed points and cusps in the heating rate for Ex.~4 in Table~\ref{Table:Examples}}
\label{App:Analytical_results:Ex4}


Consider $H_{2}$ given by the deformation of Ex.~4 in Table~\ref{Table:Examples}, i.e.,
\begin{equation}
w_{2}(\xi) = \frac{a}{b + \cos(2\pi\xi) + \sin(2\pi k \xi)}
\label{example44}
\end{equation}
for $a > 0$, $b > 2$, and $k \in \mathbb{Z}^{+}$.
As we will show, this case has the advantage of analytical tractability at the same time as the parts of the phase diagram due to only fixed points allow for interesting internal structure:
There can be cusps in the heating rate and regions with different number of unstable fixed points.

For clarity, we stress that we will consider only fixed-point ($p = 1$) solutions to \eqref{p-periodic_points} in what follows.

\begin{figure}[!htbp]

\centering

\includegraphics[scale=1, trim=0 0 0 0, clip=true]
{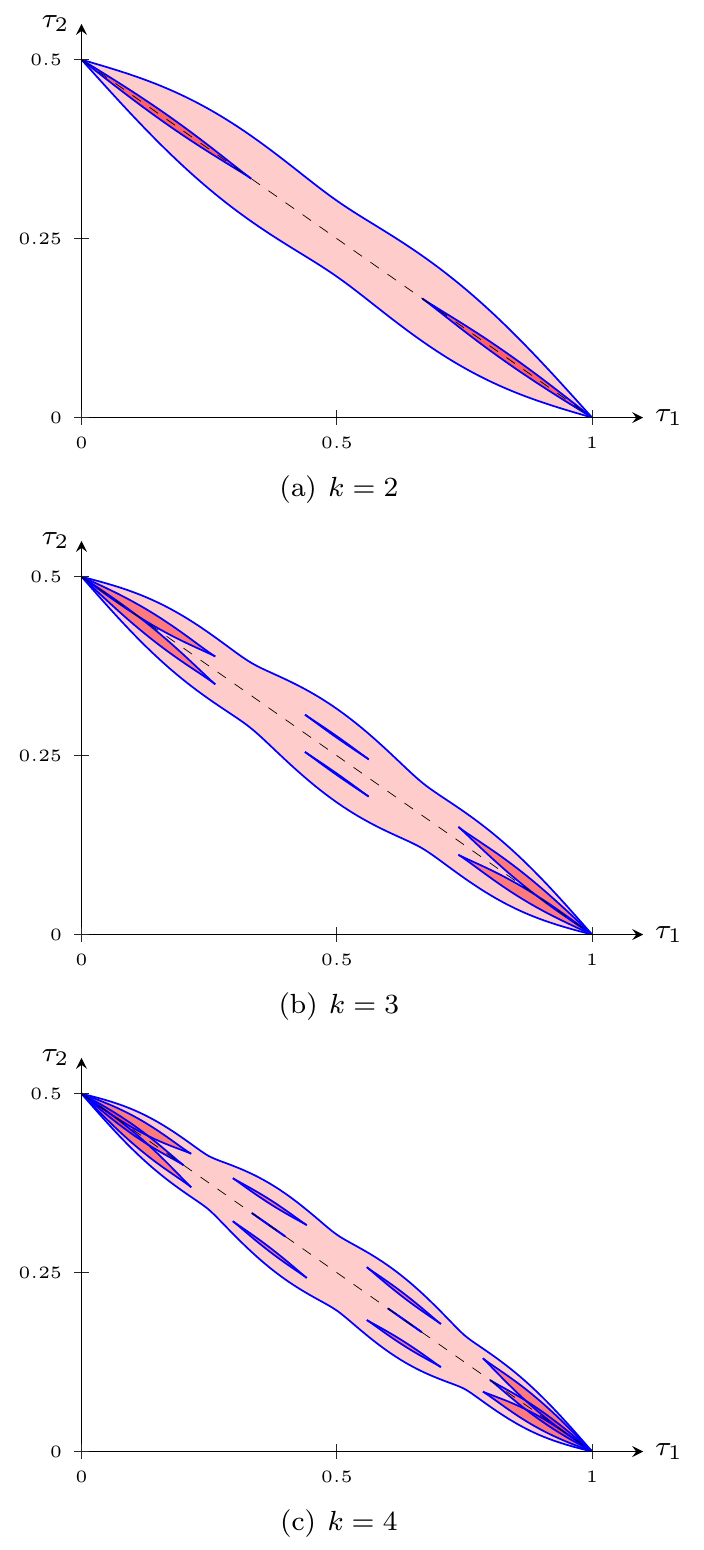}%

\vspace{-2.0mm}

\caption{%
Phase diagrams showing heating phases due to only fixed points for the special case when $H_{1}$ is homogeneous and $H_{2}$ is given by Ex.~4 with $a=6$, $b=3$, and
(a)
$k=2$,
(b)
$k=3$,
or
(c)
$k=4$.
In all diagrams, the center leaf-shaped region in light pink contains two unstable fixed points (one for each of the $\pm$-components), while leaves in dark pink contain four or more (an intersection of two leaves with $m$ unstable fixed points each yields a new leaf with $m + 2$ unstable fixed points).
See the caption to Fig.~\ref{Fig:Phase_diagrams} for further details.%
}

\label{Fig:Phase_diagrams_fp_SinCos_CFT}

\end{figure}

First, we compute the transition lines between regions with different number of unstable fixed points.
To do so, we note that \eqref{tau_1_xi_c_special} can be written as
\begin{equation}
\sin( \pi k \tau_{1} ) \cos( 2\pi k \xi_{\mathrm{c}} )
= \sin( \pi \tau_{1} ) \sin( 2\pi \xi_{\mathrm{c}} ),
\end{equation}
which has $2k$ solutions.
Inserting these into \eqref{tau_2_xi_ast_special} gives the transition lines.
The obtained heating phases due to only fixed points are shown in Fig.~\ref{Fig:Phase_diagrams_fp_SinCos_CFT} for $a = 6$, $b = 3$, and different $k$.
One can observe a simple pattern:
If $k$ increases by one, the number of leaf-shaped regions at the extremities of the phase diagram increases by one.
Such leaves correspond to regions with four unstable fixed points (two for each component).
Whenever these leaves intersect, their intersections have six unstable fixed points, and higher-order intersections will have more unstable fixed points.
In the middle of the overall heating phase, some leaves also appear, but they do not intersect and therefore have four unstable fixed points each.

\begin{figure}[!htbp]

\centering

\includegraphics[scale=1, trim=0 0 0 0, clip=true]
{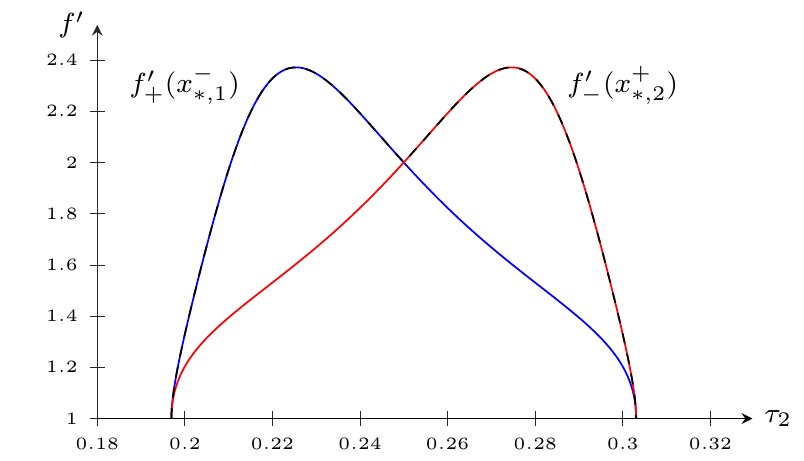}%

\vspace{-2.0mm}

\caption{%
Plots of $f_{+}'(\tx_{\ast, 1}^{-})$ (blue curve) and $f_{-}'(\tx_{\ast, 2}^{+})$ (red curve) obtained using \eqref{xi_ast_j_Ex4} as a function of $\tau_{2}$ for $\tau_{1} = 1/2$ fixed.
The black dashed curve is the maximum, corresponding to the heating rate through \eqref{heating_rate}, and has a cusp at $\tau_{2} = 1/4$.%
}

\label{Fig:Heating_rate_cusp}

\end{figure}

Second, we analytically show the existence of a cusp in the heating rate for the deformation in \eqref{example44} with $a = 6$, $b = 3$, and $k = 2$.
We focus on the large center leaf in Fig.~\ref{Fig:SinCos_CFT_1}(a) [equivalently, Fig.~\ref{Fig:Phase_diagrams_fp_SinCos_CFT}(a)] due to fixed points only:
The heating rate is then obtained by finding all $\xi_{\ast}$ solving \eqref{tau_2_xi_ast_special}, computing the corresponding $\tx_{\ast}^{\mp}$, inserting these into \eqref{f_deriv_pm_x_mp_ast} to obtain the derivatives, and finally inserting the latter into \eqref{heating_rate}.
For our particular choice of deformation, finding fixed points is equivalent to solving
\begin{multline}
\label{fp_eq_Ex4}
2\sin(\pi\tau_1) \cos(2\pi \xi_{\ast})
+ \sin(2\pi\tau_1) \sin(4\pi \xi_{\ast}) \\
= -12 \pi \left( \tau_{2} + \frac{\tau_{1}}{2} \right).
\end{multline}
In this case, the periodicities in the $\tau_{1}$- and $\tau_{2}$-directions are $1$ and  $1/2$, respectively.
Thus, the center of the phase diagram in Fig.~\ref{Fig:Phase_diagrams_fp_SinCos_CFT}(a) is $(\tau_{1}, \tau_{2}) = (1/2, 1/4)$,
and Fig.~\ref{Fig:SinCos_CFT_1}(c) indicates that it corresponds to a cusp in the heating rate.
Note that $(\tau_{1}, \tau_{2}) = (1/2, 1/4)$ lies in a heating phase with two unstable fixed points $\tx_{\ast, 1}^{-}$ and $\tx_{\ast, 2}^{+}$, one for each of the $\pm$-components, and is equivalent to $(1/2, -1/4)$ due to the symmetries illustrated in Fig.~\ref{Fig:Phase_diagrams}.
To interpret the equations correctly, we use $\tau_{2} \in [-1/2, 0]$ in practice:
Solving \eqref{fp_eq_Ex4} for $(\tau_{1}, \tau_{2}) = (1/2, -1/4)$ yields two solutions $\xi_{\ast}^{1} = 1/4$ and $\xi_{\ast}^{2} = -1/4$, and it follows from
$\tx_{\ast, j}^{\mp} = L(\xi_{\ast}^{j} \pm 1/4)$
[cf.\ \eqref{x_mp_ast_xi_ast}] and \eqref{f_deriv_pm_x_mp_ast} that $\tx_{\ast, 1}^{-} = L/2$ and $\tx_{\ast, 2}^{+} = - L/2$ with
$f_{+}'(\tx_{\ast, 1}^{-}) = 2 = f_{-}'(\tx_{\ast, 2}^{+})$,
i.e., both unstable fixed points yield the same contribution to the heating rate, see Fig.~\ref{Fig:Heating_rate_cusp}.

To prove that $(\tau_{1}, \tau_{2}) = (1/2, 1/4)$ corresponds to a cusp in the heating rate, we set $\tau_1 = 1/2$ and study the fixed points as functions of $\tau_{2}$.
The solutions to \eqref{fp_eq_Ex4} are now (recall that we use $\tau_{2} \in [-1/2, 0]$ in practice):
\begin{equation}
\label{xi_ast_j_Ex4}
\begin{aligned}
\xi_{\ast}^{1}
& = + ({1}/{2\pi}) \arccos(-3\pi [1 + 4\tau_{2}]/2), \\
\xi_{\ast}^{2}
& = - ({1}/{2\pi}) \arccos(-3\pi [1 + 4\tau_{2}]/2).
\end{aligned}
\end{equation}
The corresponding $f_{+}'(\tx_{\ast, 1}^{-})$ and $f_{-}'(\tx_{\ast, 2}^{+})$ are plotted in Fig.~\ref{Fig:Heating_rate_cusp} as functions of $\tau_{2}$.
Using \eqref{f_deriv_pm_x_mp_ast}, we obtain
\begin{equation}
\begin{aligned}
\lim_{\tau_2 \to -1/4}
\partial_{\tau_2}
f_{+}'(\tx_{\ast, 1}^{-})
& = -6\pi, \\
\lim_{\tau_2 \to -1/4}
\partial_{\tau_2}
f_{-}'(\tx_{\ast, 2}^{+})
& = + 6\pi,
\end{aligned}
\end{equation}
implying that the maximum of $f_{+}'(\tx_{\ast, 1}^{-})$ and $f_{-}'(\tx_{\ast, 2}^{+})$ 
is not differentiable at $(\tau_{1}, \tau_{2}) = (1/2, 1/4)$.
This together with \eqref{heating_rate} proves the existence of a cusp in the heating rate, and thus that this is a possibility in general.


\section{Computation of entanglement entropy}
\label{App:EE}


In what follows, we compute the von Neumann entanglement entropy $S_{A}(t)$ for a subsystem on the interval $A = [x, y]$ with the rest of the system.
This can be defined as
\begin{equation}
S_{A}(t)
= \lim_{m \to 1}
	\frac{1}{1-m}
	\log \Bigl( \Tr \bigl[ \hat{\rho}_{A}(t)^{m} \bigr] \Bigr),
\end{equation}
where $\hat{\rho}_{A}(t) = U_{F}^{n} \hat{\rho}_{A} U_{F}^{-n}$ for $t = n t_{\mathrm{cyc}}$ and $\hat{\rho}_{A}$ is the reduced density matrix for the subsystem on $A$.

The computation of $S_{A}(t)$ can be done with the help of twist fields $\Phi_{m}(x; t)$, whose conformal weights are $\Delta^{\pm}_{m} = (c/24)(m - 1/m)$ \cite{CCaD:2008}:
\begin{equation}
S_{A}(t)
= \lim_{m \to 1}
	\frac{1}{1-m}
	\log \bigl[ \langle0| \Phi_{m}(x; t) \Phi_{m}(y; t) |0\rangle \bigr],
\end{equation}
see \cite{CaCa2:2016} for a review.
Using the result in \eqref{Phi_2pcf} and taking the limit yields \eqref{EE}.

We conclude with two remarks.

\begin{figure}[!htbp]

\centering

\includegraphics[scale=1, trim=0 0 0 0, clip=true]
{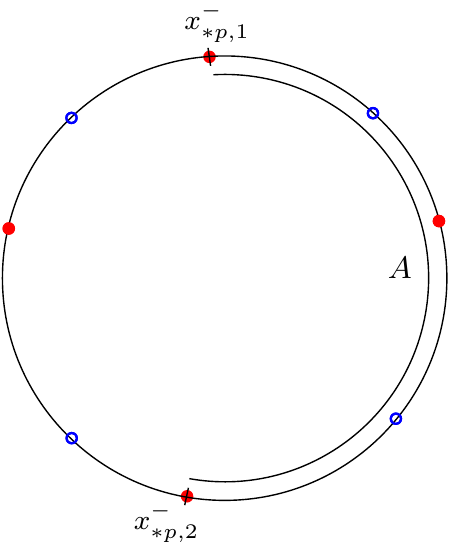}%

\vspace{-2.0mm}

\caption{%
Illustration of the additional case (c) described in the text [cf.\ \eqref{S_pm_case_c}] for the entanglement entropy $S_{A}(t)$ in the heating phase.
See the caption to Fig.~\ref{Fig:Entanglement_pattern_Case_ab} for further details.%
}

\label{Fig:Entanglement_pattern_Case_c}

\end{figure}

First, for the more general case where the two components in the Hamiltonian have different drives, we note that the entanglement entropy can also decay linearly in time, different from cases (a) and (b) in Sec.~\ref{Sec:EE}.
For example, this can happen in the following case:%

(c)
Suppose that there are at least two unstable periodic points $\tx_{\ast p, 1}^{-}$ and $\tx_{\ast p, 2}^{-}$ with respect to the right-moving component, while the left-moving component remains undriven (or vice versa), and that the interval is chosen as
$A = [\tx_{\ast p, 1}^{-}, \tx_{\ast p, 2}^{-}]$, see Fig.~\ref{Fig:Entanglement_pattern_Case_c}.
In this case,
\begin{align}
\label{S_pm_case_c}
S_{+}(t)
& = - \frac{
		\log \bigl[ f_{+}^{p \,\prime}(\tx_{\ast p, 1}^{-}) \bigr]
		+ \log \bigl[ f_{+}^{p \,\prime}(\tx_{\ast p, 2}^{-}) \bigr]
	  }{p t_{\mathrm{cyc}}} \, t + o(t), \nonumber \\
S_{-}(t)
& = o(t).
\end{align}
Since $f_{+}^{p \,\prime}(\tx_{\ast p, 1}^{-})$ and $f_{+}^{p \,\prime}(\tx_{\ast p, 2}^{-}) > 1$, the entanglement entropy decays linearly in time.
We stress that the system is still heating, i.e., the total energy grows exponentially even though the entanglement entropy decreases.
This generalizes previous results in \cite{WFVG1:2020} for Floquet drives in deformed CFTs where the Hamiltonian is a linear combination of the Virasoro generators $\{ L^{\pm}_{0}, L^{\pm}_{m}, L^{\pm}_{-m} \}$ for some $m \in \mathbb{Z}^{+}$ \cite{Note1} to inhomogeneous CFT with general smooth deformations.

Second, if there is only one unstable fixed point per component (by definition, there cannot be only one periodic point with period $p > 1$, if such exists), the entanglement entropy cannot grow for long times in the case of periodic boundary conditions (which we have here).
However, using open boundary conditions, the right- and left- moving components are connected and the entanglement entropy grows linearly in time if $A$ contains one of the two unstable fixed points, as discussed for the special case of $\mathfrak{sl}(2)$-deformations in \cite{WeWu2:2018, FGVW:2020}, where the deformed Hamiltonian is a linear combination of the Virasoro generators $\{ L^{\pm}_{0}, L^{\pm}_{1}, L^{\pm}_{-1} \}$ \cite{Note1}.


\section{Computation of mutual information}
\label{App:MI}


Below we compute the mutual information
$I_{A; B}(t) = S_{A}(t) + S_{B}(t) - S_{A \cup B}(t)$
for two intervals $A = [x_1, x_2]$ and $B = [x_3, x_4]$.
We give details only for the right-moving component (without loss of generality, since the two $\pm$-components commute).
Moreover, it follows from Sec.~\ref{Sec:EE} that the result is nontrivial if and only if $A$ and $B$ contain at least one unstable periodic point each, denoted $\tx_{\ast p, A}^{-}$ and $\tx_{\ast p, B}^{-} \neq \tx_{\ast p, A}^{-}$, respectively,
see Fig.~\ref{Fig:Mutual_info}; we will assume this in what follows.

Computing $S_{A}(t)$ and $S_{B}(t)$ can be done using \eqref{EE}.
To compute $S_{A \cup B}(t)$ requires a four-point correlation function of twist fields (cf.\ Appendix~\ref{App:EE}) of the form
\begin{align}
\label{S_A_cup_B_t_4pcf}
& \Biggl(
\prod_{j = 1}^{4}
\biggl[ \frac{\partial \tx_{t}^{-}(x_{j})}{\partial x} \biggr]^{\Delta^{+}_{m}}
\Biggr) \\
& \quad \times
\biggl(
	\frac{\ii\pi}{L\sin(\pi[\tx_{t}^{-}(x_{1}) - \tx_{t}^{-}(x_{4}) + \ii 0^+]/L)}
\biggr)^{2\Delta^{+}_{m}} \nonumber \\
& \quad \times
\biggl(
	\frac{\ii\pi}{L\sin(\pi[\tx_{t}^{-}(x_{2}) - \tx_{t}^{-}(x_{3}) + \ii 0^+]/L)}
\biggr)^{2\Delta^{+}_{m}}
\mathcal{F}(u), \nonumber
\end{align}
where $\mathcal{F}(u)$ is the conformal block, which depends on the operator content of the theory, and
\begin{align}
u
& = \frac{
		\sin(\pi[\tx_{t}^{-}(x_1) - \tx_{t}^{-}(x_4) + \ii 0^+]/L)
	}{
		\sin(\pi[\tx_{t}^{-}(x_1) - \tx_{t}^{-}(x_2) + \ii 0^+]/L)
	} \nonumber \\
& \quad \times
	\frac{
		\sin(\pi[\tx_{t}^{-}(x_2) - \tx_{t}^{-}(x_3) + \ii 0^+]/L)
	}{
		\sin(\pi[\tx_{t}^{-}(x_4) - \tx_{t}^{-}(x_3) + \ii 0^+]/L)
	}
\end{align}
is the so-called cross ratio.
We will use that the conformal block $\mathcal{F}(u)$ does not contribute if $u$ tends to a constant finite value different from $1$, see, e.g., Appendix B of \cite{FGVW:2020}.

For the two cases (a) and (b) described in Sec.~\ref{Sec:MI} the following holds:

(a)
Since $\tx_{t}^{-}(x_2)$ and $\tx_{t}^{-}(x_3)$ flow to $\tx_{\ast p, AB}^{-}$, while $\tx_{t}^{-}(x_1)$ and $\tx_{t}^{-}(x_4)$ flow to two other different stable periodic points $\tx_{\ast p, 1}^{-}$ and $\tx_{\ast p, 4}^{-}$, it follows that $u$ tends to $0$, implying that $\mathcal{F}(u)$ does not contribute for large $t$ and can be neglected in what follows.
Due to Property~\ref{Item:Convergence} in Sec.~\ref{Sec:Main_tools:Properties_pps}, the contributions to the von Neumann entanglement entropy $S_{A \cup B}(t)$ from $\tx_{t}^{-}(x_2)$ and $\tx_{t}^{-}(x_3)$ cancel [see the argument for the case (b) in Sec.~\ref{Sec:EE}], and thus the leading contribution from the right-moving component to $S_{A \cup B}(t)$ is
\begin{multline}
\label{S_A_cup_B_t_lead_p_contrib}
- \frac{c}{12}
	\log \Biggl[
		\frac{\partial \tx_{t}^{-}(x_1)}{\partial x}
		\frac{\partial \tx_{t}^{-}(x_4)}{\partial x} \\
\times
		\biggl(
			\frac{\ii\pi}{L\sin(\pi[\tx_{t}^{-}(x_1) - \tx_{t}^{-}(x_4) + \ii 0^+]/L)}
		\biggr)^{2}
	\Biggr].
\end{multline}
The derivative factors above decay exponentially as $\tx_{t}^{-}(x_1)$ and $\tx_{t}^{-}(x_4)$ flow to their respective stable periodic points, which implies that \eqref{S_A_cup_B_t_lead_p_contrib} goes as
\begin{equation}
- \frac{c}{12}
	\frac{
		\log[f_{+}^{p \,\prime}(\tx_{\ast p, 1}^{-})]
		+ \log[f_{+}^{p \,\prime}(\tx_{\ast p, 4}^{-})]
	}{p t_{\mathrm{cyc}}} \, t + o(t)
\end{equation}
for large $t$.
This and the results for $S_{A}(t)$ and $S_{B}(t)$ [cf.\ \eqref{cS_EE_1upp_lt}] imply that the leading contribution to $I_{A; B}(t)$ from the right-moving component is given by \eqref{MI_rmc_case_a}.

(b)
We can repeat the argument in (a), recalling that $\tx_{t}^{-}(x_{j})$ ($j = 1, 2, 3, 4$) flow to four different stable periodic points.
Again, $u$ tends to a constant different from $1$, and thus $\cF(u)$ does not contribute for large $t$, as before.
The derivative terms in \eqref{S_A_cup_B_t_4pcf} will all decay exponentially, while the denominators stay finite.
It follows that the leading contribution to $S_{A \cup B}(t)$ is canceled by the corresponding ones to $S_{A}(t) + S_{B}(t)$, meaning that the contribution from the right-moving component to $I_{A; B}(t)$ does not change with time for large $t$.

To summarize, the intervals $A$ and $B$ have to contain at least one unstable periodic point each and to be separated by a single stable periodic point for $I_{A;B}(t)$ to grow linearly in time for large $t$.


\end{appendix}



\end{document}